\documentclass[11pt]{JHEP3} 
\pdfoutput=1
\usepackage{graphicx}

\newcommand{\be}{\begin{equation}}
\newcommand{\ee}{\end{equation}}
\newcommand{\bea}{\begin{eqnarray}}
\newcommand{\eea}{\end{eqnarray}}
\newcommand{\bary}{\begin{array}}
\newcommand{\eary}{\end{array}}

\newcommand{\nn} {\nonumber \\}

\newcommand {\eqr} [1]  {{(eq.\,\ref{eq:#1})}}

\newcommand{\file}[1]{}

\newcommand{\pd}    {\partial}

\title{Galaxy rotation curves from string theory} 

\author{
Yeuk-Kwan E. Cheung\\
 Department of Physics, Nanjing University\\
22 Hankou Road, Nanjing 210093, China\\ 
E-mail: \email{cheung(at)nju.edu.cn}
}

\author{
Hsien-Chung Kao\thanks{on leave from National Taiwan Normal University}\\
Department of Mathematics, University of Durham\\
Durham, DH1 3LE, United Kingdom
}

\author{
Konstantin Savvidy\\ 
Department of Physics, Nanjing University\\
22 Hankou Road, Nanjing 210093, China
}

\abstract{This is a speculative attempt to connect string theory with cosmological observation.   
Inspired by an exactly solvable model in string theory, and based on the assumption that all matter is made of strings, 
individual stars will couple universally to this  string gauge field and execute Landau orbits, much like electrons in  an external magnetic field.
This  three-parameter phenomenological  model can adequately fit the galaxy rotation curves.
The extra centripetal acceleration provided by the background field can  hence  account 
for the ``missing mass'' needed to sustain the high  rotation speed beyond the bulk  of the stellar mass. 
The rotation speed of the stars on the outskirts  of a  galaxy is predicted to be  linearly rising with distance.
 } 

\preprint{arXiv:astro-ph/0702290}

\begin{document}

Dark Matter has been accepted by many  as the leading candidate to explain the ``missing mass'' problem in  galaxies.  The problem arises because the rotational velocities of the stars about  the center of the galaxy  does  not fall off in the Keplerian way as one exits the stellar disk.  
The stellar disk accommodates most of the young and bright stars which in turn  are supposed to account for  the lion's share of the mass in a typical spiral  galaxy.    Instead, the rotational speed of the stars stays flat up to a distance ten times the size of the galactic disk.   In order to explain the conundrum,  Dark Matter is proposed to fill up the void in the form of a gigantic spherical halo (in the simplest case)  engulfing the visible part of the galaxy in such a way that the rotational speed of the stars would be kept constant in the desired region.    Usually in order to fit the data the dark matter halo alone is modelled with three parameters.  Here we would like to entertain the possibility that a string background field that couples uniquely to the worldsheet of the strings, 
{\em{i.e.}}  extremely natural from the point of view of string theory, can take up the bill of the missing mass and reproduce the galaxy rotation curves.  

There are many  contenders  to the Dark Matter proposal with various degrees of success.
%MOND,  an attempt to modify Newtonian gravity at large distances~\cite{Milgrom:1983pn} (and  reviewed in  \cite{Sanders:2002pf}),  has  been suggested as a possible replacement  of  Dark Matter.    
We, instead,  like to show that  if one accepts the assumption that all matter is made of strings, these fields that arise naturally from string theory can put the galaxy into Landau orbits, much like an electron in the presence of a magnetic field.  This produces an  extra centripetal acceleration that would have been otherwise attributed to the presence of extra mass in the galaxy.  Using a  three parameter model we can already capture the rotational curves of galaxies of different sizes and shapes well over the extent of the galactic disk.  With a suitable choice of parameters the rotational speed can be stabilized up to twenty times  the  radius of the stellar disk, $R_{d}$.   We  perform the fitting on twenty-two spiral galaxies with a sizeable range in luminosity, as well as size,  and in different Hubble types. We obtain  an  average $\chi$-squared  of $1.60$ with a  standard deviation of $1.68$,  indicating very good fitting.  A macroscopic model that describes this Landau rotation as well as gravitational lensing  has been  constructed from string sigma model and will be presented  in  forthcoming articles.

Twenty-two   galaxies
%~\footnote{One galaxy turns out to be two merging galaxies.  It is dropped from the subsequent analysis after the rotation curve fitting.} 
have been fit to the model without allowing for {{\it any}} Dark Matter components.  This is done to test the limits of our model.  In reality there are many well-motivated candidates for Dark Matter which we believe will be detected in future observations.  There are a wide range of astronomical objects not giving light in the visible spectrum and therefore not included in the estimate of  the mass of the galaxy from luminosity.  They can be  stars that are not heavy enough to ignite hydrogen, or black holes which take in all the light that shines on them.  
We should allow for the possibility of such ordinary matter being
detected directly as a result of advances in technology. The other more
likely possibility in the light of the measurements by WMAP~\cite{wmap}  is that
the Dark Matter is non-baryonic and consists of an exotic neutral
particle such as the supersymmetric partners or other new candidates to
be discovered in the new generation of  high energy experiments.
The conundrum involving  Dark Energy and Dark Matter may also be resolved due to theoretical advances and we are back to the familiar universe.   We may not like such an ordinary universe, without any inexplicable components.  But we should keep in mind that there is such a possibility.  
  
\section{The model}

In String Theory~\cite{stringtheory} it is postulated that the most fundamental constituent of matter is a string, a one-dimensional object.
\begin{figure}[h!!]
%\begin{center}
    \includegraphics[scale=0.5]{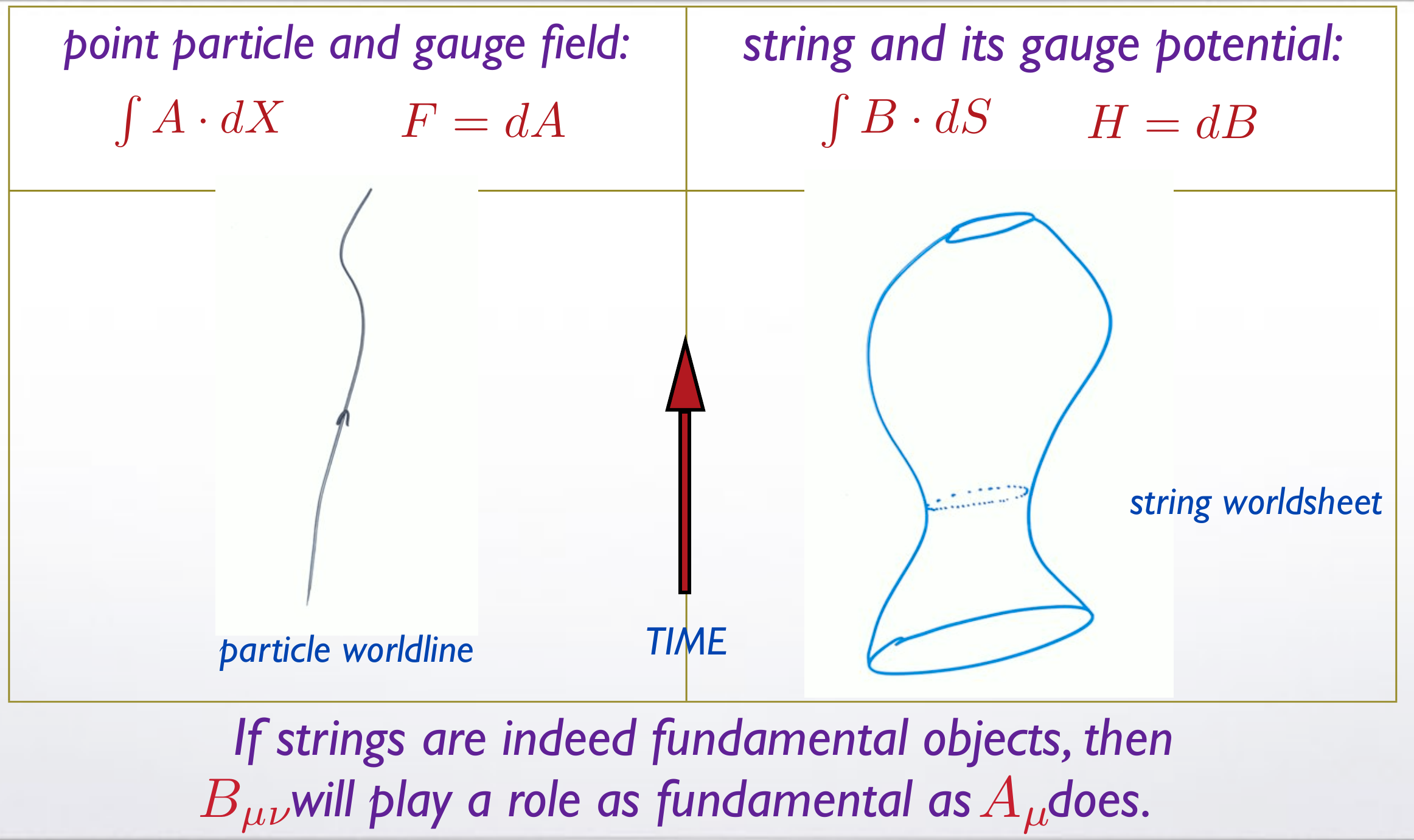}
%\end{center}
\caption{\small Particle worldline and string worldsheet and their  couplings to the corresponding gauge potentials.}
 \label{fig:worldsheet}
 %\vspace{-0.1in}
\end{figure}
The observed particle spectrum corresponds to different quanta of excitations of the string.    In particular the massless spectrum consists of a scalar, gauge bosons, graviton and their fermionic partners in a supersymmetric string theory.  In one stroke it unifies Higgs-like scalars (spin zero), gauge bosons  (spin one)  and graviton (spin two), together with their fermionic partners of half-integral spins,  in one framework.    The gauge theory of Yang-Mills fields that governed the interaction of point  particles  is extended to a gauge theory of a two-form gauge potential.  This gauge potential couples uniquely to the  worldsheet of the strings.  
See Fig.~\ref{fig:worldsheet}.

In one type of string background,  corresponding to plane-polarized gravitational fields, string theory can be solved  exactly~\cite{pp-waves}.    
Furthermore in one particular model, the Nappi-Witten model~\cite{Nappi:1993ie}, the space where the strings propagate  is as close to four-dimensional Minkowski space as one can get with the presence of the string background field\footnote{As soon as the background gauge field is turned on, the spacetime back reacts and settles into another spacetime with a different geomety.  In other words one cannot treat the resultant spacetime as  the Minkowski space with a small perturbation due to the field.}.   
This model  enjoys an  infinite dimensional symmetry at the quantum level~\cite{Nappi:1993ie} which in turn enables us to  obtain a  complete and covariant 
 solution~\cite{Cheung:2003ym}  of the first quantized string theory in this background\footnote{%.  
 The construction of  the string vertex operators that are responsible for the creation of the string states, and the computation of  the correlation functions of an arbitrary number of scattering particles have been reported in~\cite{Cheung:2003ym}.}.  This model is valid at all energy scales as long as the string coupling is small.  The symmetry of the model is exactly the same at high energy as at low energy.  This in turn lends hope that the certain  phenomenon which originates at higher energy scale  from string theory  will persist at an energy much lower and a scale much  bigger than those natural for  the fundamental strings.    
The corresponding  string  sigma model action is 
 \begin{equation}
L\sim \int_{\Sigma} G_{\mu\nu}\, \pd X^{\mu} \cdot \pd X^{\nu} + B_{\mu\nu}\,  dX^{\mu}\wedge\, dX^{\nu} 
\end{equation}
where $X(\sigma)$ is a mapping  from the  two-dimensional worldsheet, denoted by $\Sigma$,  to the  space-time, parametrized by $X$'s with a metric $G_{\mu\nu}$,    in which strings  can propagate and interact.  
It is this two-dimensional conformal theory with quantum fields, $X$'s (and possibly $\Psi$'s in the supersymmetric version) capable of capturing the interaction of the strings and hence offers a UV finite completion of Einstein's theory of general relativity. 
 
The geometry of the plane-polarized gravitational wave background is encoded in the metric  $G_{\mu\nu}$:
 \begin{eqnarray}
  \label{eq:Gmunu}
 G_{mn}&=& \pmatrix{0&1 & a_2H & -a_1H\cr 
 							1 &0 &  0  &0\cr
							a_2H&  0 & 1 & 0  \cr
							 -a_1H &0& 0 & 1}~.
 \end{eqnarray}
\begin{figure}[h!!]
%\begin{center}
    \includegraphics[height=2.0in]{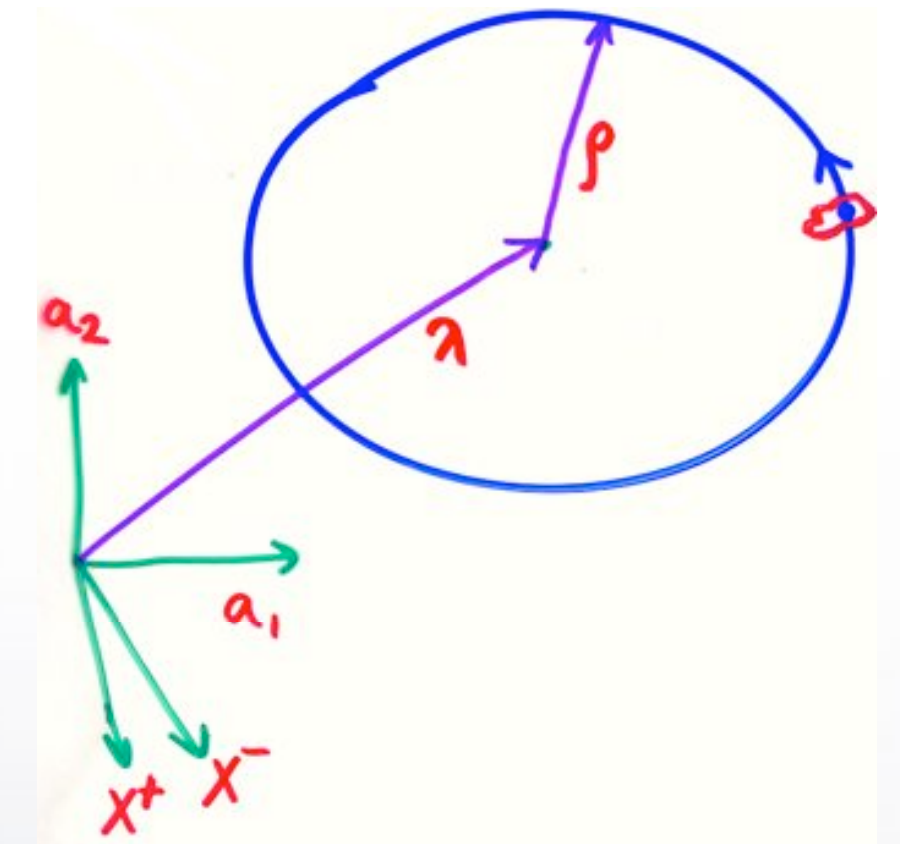}
%\end{center}
\caption{\small Closed strings execute circular rotation in the presence of the gravimagnetic fields with the center of mass lying on the Landau orbits.}
 \label{fig:landau}
 %\vspace{-0.1in}
\end{figure}
The gauge potential, $B_{\mu\nu}$, coupled to the worldsheet of strings,  is linear,  $B=H\cdot X$,  giving rise to a constant field strength, $H$: 
 \begin{eqnarray}
  \label{eq:Bmunu}
 B_{mn} &=& \pmatrix{   0&1 & a_2H & -a_1H\cr
 								1 &0 &  0  &0\cr
                 					-a_2H&  0 & 1 & 0  \cr
								 a_1H &0& 0 & 1 
				}~.
\end{eqnarray}
In the presence of this field, the 
center of mass of the closed 
strings undergoes  circular motion  in the $(a_{1},a_{2})$ plane transverse to the light cone, $(X^{+}, X^{-})$, much like  the motion of electrons in the presence of a constant magnetic field perpendicular to the plane of rotation.  The centre of mass of these strings  follows  the Landau orbits as depicted  in  Fig.~\ref{fig:landau}.

If all matter is indeed made of strings each star inside the galaxy can then  be thought of a coherent state of its constituent strings.  The  whole galaxy will be charged under this gravimagnetic field and each  star will likewise follow the same trajectory as the individual string  constituents\footnote{In a forthcoming article we formulate an effective field theory description for this process.}.   Therefore the equation of motion of an arbitrary   star in the galaxy will be modified by presence of this additional Lorentz-like force:
\be 
\frac{v^2}{r}  = \frac{Q}{m} H v +  F_{N}(\vec{r})
\ee
where $F_{N}(\vec{r})$ is the Newtonian attraction due to a stellar mass distribution $M(r)$  within a radius of  $r$.   $H$ is the magnitude of the gravimagnetic field perpendicular to the galactic plane while $Q$ is the charge of the stellar matter under this field.  Note $Q$ gives rise to the gravitational attraction.  So it should be proportional to $m$ as we shall see.  
To  model the distribution  of the stars in the stellar disk  we take a continuum limit and use the parametric distribution with exponential falloff due to van der Kruit and Searle:
\be       \label{eq:KS}
\rho_{KS}(r, z) = \rho_{0}\,  \textrm{exp}\,(-\frac{r}{R_{d}})\,\,  \textrm{sech}^{2}\, (\frac{6z}{R_{d}})
\ee
with $\rho_{0}$ being the central matter density.  $R_{d}$ is the characteristic radius of a galaxy, and $Z_{d}$ the characteristic thickness.
Let us stress that   $\rho_{KS}$  is the mass due to the  $visible$ stars only.  There is no allowance for contribution from the Dark Matter halo.  We have also ignored the contribution from gas and dust.

We first compute the force at radius ${r}$ due to the  mass distribution:
\bea   \label{eq:force}
 F_{N} (r)  &=& - G_{N}\, \nabla_{\vec{r}}\, \int \frac{ \rho_{KS} (\vec{r}\prime) } {|\vec{r} - \vec{r}\prime|} \,r'\, dr'\, dz\, d\phi       \nn 
       &=& \rho_{0} \, G_{N}\,  R_{d}\,  \tilde{F}(\tilde{r})~.
\eea 
where the integration is taken over all spatial volume.    
In the last  line we have integrated over $z$ and $\phi$ and defined  a dimensionless radius parameter   $\tilde{r} \equiv \frac{r}{R_{d}}$.
Hence $\tilde{F}(\tilde{r})$, a dimensionless quantity,  can  be thought of as the universal Newtonian force function due to a vdK-S mass distribution and can be computed numerically.  
Two additional parameters,  $\Omega\equiv \frac{QH}{2m}$, and $\rho$, are introduced into our model as follows:
\be  \label{eq:model}
v^{2} -  2 \Omega \,r \,  v  - \rho \, r\, \tilde{F}(\tilde{r}) = 0 
\ee
whereas $\rho$ is related to the original $\rho_{0}$ by:
\be   \label{eq:rho-rho0}
 \rho = \rho_{0} \, G_{N}\,  R_{d}.
 \ee
Of the parameters in the model $\rho$ and $ R_{d}$ can be crosschecked or fit with the photometric data using the vdK-S formula, whereas $m$, $Q$ and therefore $\Omega$ are not available through other observations as yet.  
%Nevertheless $\Omega$ can be related to the parameters of the MOND model.

\section{A falsifiable prediction}

Let us now examine the physical consequencies due to the addition of a term linear in $v$ into the force law~\eqr{force}.
At radius smaller than $R_{d}$, the rotational velocity rises linearly as it is in the case of pure Keplerian way, except with a steeper slope due to the introduction of the background rotation.  
So it will attain the maximum rotation speed faster at a smaller radius, given a mass distribution.  This means that less mass is needed to reach  a given rotation speed.  Therefore we expect that the total mass of a galaxy determined from this model will be smaller than the one from  a  purely  Newtonian model.  Furthermore  those galaxies which enjoy the background rotation have  much more reasonable mass to light ratios, in the vicinity of unity (as tabulated in the last column of Table~\ref{table:bestfit} and shown in  Figure~\ref{fig:mass-to-light}  in Section~\ref{sec:data} below) than the ones would have gotten from the Dark Matter model.

As we exit the galactic disk where the majority of the stars reside, at a distance  exceeding  $2\,R_{d}$, a drop in the Newtonian attraction due to the decrease in  stellar mass is compensated by the linear rise in $v\sim \Omega r$ such that a plateau in rotation speed  is obtained.   By tuning  $\Omega$ in the model one can generically obtain a plateau extending from  $2 R_{d}$  to about $20R_{d}$.    Way beyond the extent of the stellar disk the background rotation dominates and the rotation curve rises linearly again.  This prediction is compatible with the rising rotation velocity of the individual stars on the outskirts  of a  galaxy.
This can happen at a distance  as far as $20R_{d}$ which has yet   reached by observations.  
Measurement is difficult there because of the lack of bright stars  so far away from the centre of  the galaxies.  
% We are looking for extended galaxies with rotation curve
% measurements at that distance to test our theory.  

This linear  rise in  rotation speed far away  from the centre of the galaxy  is a signature prediction of 
our model which can  be tested in the following ways:
\begin{itemize}
\item{by a larger sample of data with measurements extending to the outer part of the galaxies;}
\item{by  observations  measuring the rotation speed of small satellite galaxies;}
\item{by improved precision on current measurements.}  
\end{itemize}
If all of these prefer a flat, or even falling rotation curve in the range beyond $2.2R_{d}$, then our model is disfavoured.  
The model in the simplest form involves a constant string background field can be easily refuted by 
a direct measurement of the  falling rotation speed of satellite galaxies.  
In the future as the precision of velocity measurement around $10R_{d}$  increases one can readily distinguish the three types of models, namely the dark matter model, MOND and our model as they predicts a falling rotation curve, a flat rotation curve and a rising rotation curve respectively.  

\section{Data processing and fitting:} 
\label{sec:data}
We obtained our data of the  twenty-three  galaxies  in the SING group from FaNTOmM~\cite{data}.  
The fitted galaxy rotation curves are  presented in Appendix~A.  The stars with error bars indicate the data points while the dots indicate theoretical predictions.   
\begin{table}[htdp]   \label{table:bestfit}
\begin{center}
\begin{tabular}{||l||c|c|c|c|c|c|c||}  \hline
galaxy&   likelihood&   $\Omega$ &   $\rho$ &   $R_{d}$ (kpc)&   total mass(Msun)&   B magnitude&   $\Upsilon$ \\ \hline
 ngc0628&   4.35&   8.872&   11738&   1.803&   3.35E+10&   -20.60&   1.24\\ \hline   
 ngc0925&   2.45&   5.077&   500&   2.473&   3.68E+09&   -20.05&   0.23\\ \hline   
 ngc2403&   4.25&   13.418&   16497&   0.518&   1.12E+09&   -19.56&   0.11\\ \hline   
 ngc2976&   0.40&   0.100&   2135&   1.867&   6.76E+09&   -18.12&   2.45\\ \hline   
 ngc3031&   5.70&   2.000&   3187&   4.389&   1.31E+11&   -21.54&   2.04\\ \hline   
 ngc3184&   0.57&   0.655&   1552&   4.685&   7.77E+10&   -19.88&   5.58\\ \hline   
 ngc3198&   0.27&   1.968&   1971&   3.522&   4.19E+10&   -20.44&   1.80\\ \hline   
 ngc3521&   0.59&   6.046&   26678&   1.479&   4.20E+10&   -21.08&   1.00\\ \hline   
 ngc3621&   0.37&   10.338&   7413&   1.084&   4.60E+09&   -20.51&   0.18\\ \hline   
 ngc3938&   1.03&   9.722&   16199&   1.240&   1.50E+10&   -20.01&   0.96\\ \hline   
 ngc4236&   0.33&   0.12&   110&   4.83&   4.42E+09&   -18.10&   1.60\\ \hline   
 ngc4321&   2.07&   5.208&   3752&   3.148&   5.70E+10&   -22.06&   0.55\\ \hline   
 ngc4536&   0.74&   1.465&   734&   5.867&   7.22E+10&   -21.79&   0.89\\ \hline   
 ngc4569&   0.66&   16.370&   12262&   1.042&   6.75E+09&   -21.10&   0.16\\ \hline   
 ngc4579&   0.69&   7.385&   5164&   3.000&   6.79E+10&   -21.68&   0.93\\ \hline   
 ngc4625&   0.55&   0.100&   1140&   1.446&   1.68E+09&   -17.63&   0.96\\ \hline   
 ngc4725&   0.27&   1.378&   1523&   6.726&   2.26E+11&   -21.76&   2.87\\ \hline   
 ngc5055&   3.52&   4.064&   24581&   1.544&   4.40E+10&   -21.20&   0.94\\ \hline   
 ngc5194&   0.81&   3.186&   10718&   1.102&   6.98E+09&   -20.51&   0.28\\ \hline   
 ngc6946&   6.03&   7.440&   4980&   1.805&   1.43E+10&   -20.89&   0.40\\ \hline   
 ngc7331&   0.52&   8.465&   18613&   1.680&   4.30E+10&   -21.58&   0.64\\ \hline  
  m81dwb&   0.13&   1.119&   15919&   0.147&   2.46E+07&   -12.5&   1.58\\ \hline 
 $<\chi^{2}>$&   1.65  & ~~~ &  ~~~~  & ~~~~ & ~~~~ &~~   &~~ \\   \hline
\end{tabular}
\end{center}  \vspace{-0.1in}
\caption{The best fit values for the parameters $R_{d}$, $\rho$, and $\Omega.$ Mass for the galaxie as well as the mass to light ratio, $\Upsilon$ as computed from the best fit parameters. }
\end{table}%
The values of the best fit parameter for $R_{d}$, $\Omega$, and $\rho$, 
along with the $\chi^2$-values (per degree of freedom) of the fitting,  are tabulated in Table~1.  
%\ref{fig:bestfit}.  
Out of the twenty-three galaxies one, NGC 5713,  turns out be  two merging galaxies.  We will henceforth drop this galaxy from the rest of the analysis.

The radii in the data set are quoted in $arcsec$.  We need to convert the data into distance in 
$kilo$-$parsec$.  This is done as follows.  Let us denote the angular extent of the galaxy on the detector screen by angle $\alpha$.  We  hence have the relation:
\be
\frac{r}{d} =\frac{\alpha}{360\cdot 60 \cdot 60} \cdot \, 2\pi 
\ee 
where $d$ is the distance, in $kilo$-$parsec$,  between the galaxy and us.  
Kinematic distance modulus is encoded by the parameter ``$mucin$''  in the  FaNTOmM database.  It is the difference between the apparent magnitude, $m$,  and the absolute magnitude, $M$,  of a star or a galaxy, given by the following formula:
\be \label{eq:mucin}
mucin = m -  M  = 5 log (d) + 10 
\ee
%where $d$ is the distance measured in $kilo\,parsec$.
The in-falling velocity of the Local Group onto the Virgo cluster has been corrected for in $mucin$.  
In the case that $mucin$ is not available in the database, we use ``$mup$,'' without such correction, instead,  in the distance determination,  at the expense of a small error.  In one extreme case in our sample this introduces a $10\%$ error.   Otherwise the errors are less than $5\%$,   negligible compared to the errors in velocity measurements.   In this way we convert  the radius of the galaxy from  $arcsec$  to $kilo$-$parsec$.

Next we  turn to determining the luminous  mass of the galaxies from the density parameter  $\rho$: 
\bea      
     \label{eq:mass}
m_{*}&=& \int  \rho_{KS} (r,z) \, r \, dr \, dz \, d\phi    \nn
%              &=& \rho_{0}  \int  e^{-\frac{r}{R_{d}}}\,  sech^{2} (\frac{6z}{R_{d}}) \, r\, dr \, dz \, d\phi   \nn
%                  &=& \rho_{0} R_{d}^{3}  \int  e^{-\tilde{r}}\,  sech^{2}\,(6\tilde{z})\, \tilde{r} \, d\tilde{r} \, d\tilde{z} \, d\phi  \nn 
                  &=&   2.0944 \,  \rho_{0}\,  R_{d}^{3}
\eea
where we have integrated over
$0 < \tilde{r} < \infty$,   and $-\infty < \tilde{z}< \infty$.
 The mass of a galaxy  is  finally  computed using the   best fit value of  $\rho$ for  the  galaxy  according to   the following formula:
\be      \label{eq:rho-mass}
m_{*} = \frac{2.0944}{G_{N}} \, \rho \cdot R_{d}^{2}
\ee 
making  use of the relation~\eqr{rho-rho0} between $\rho$ and $\rho_{0}$.
Newton's constant takes the value
\be
 G_{N} = 4.32\times 10^{-6}\, km^{2}s^{-2} /kpc /M_{sun}
\ee 
in the units convenient for our calculation.

\begin{figure}[h!!]
%\begin{center}
    \includegraphics[height=2.0in]{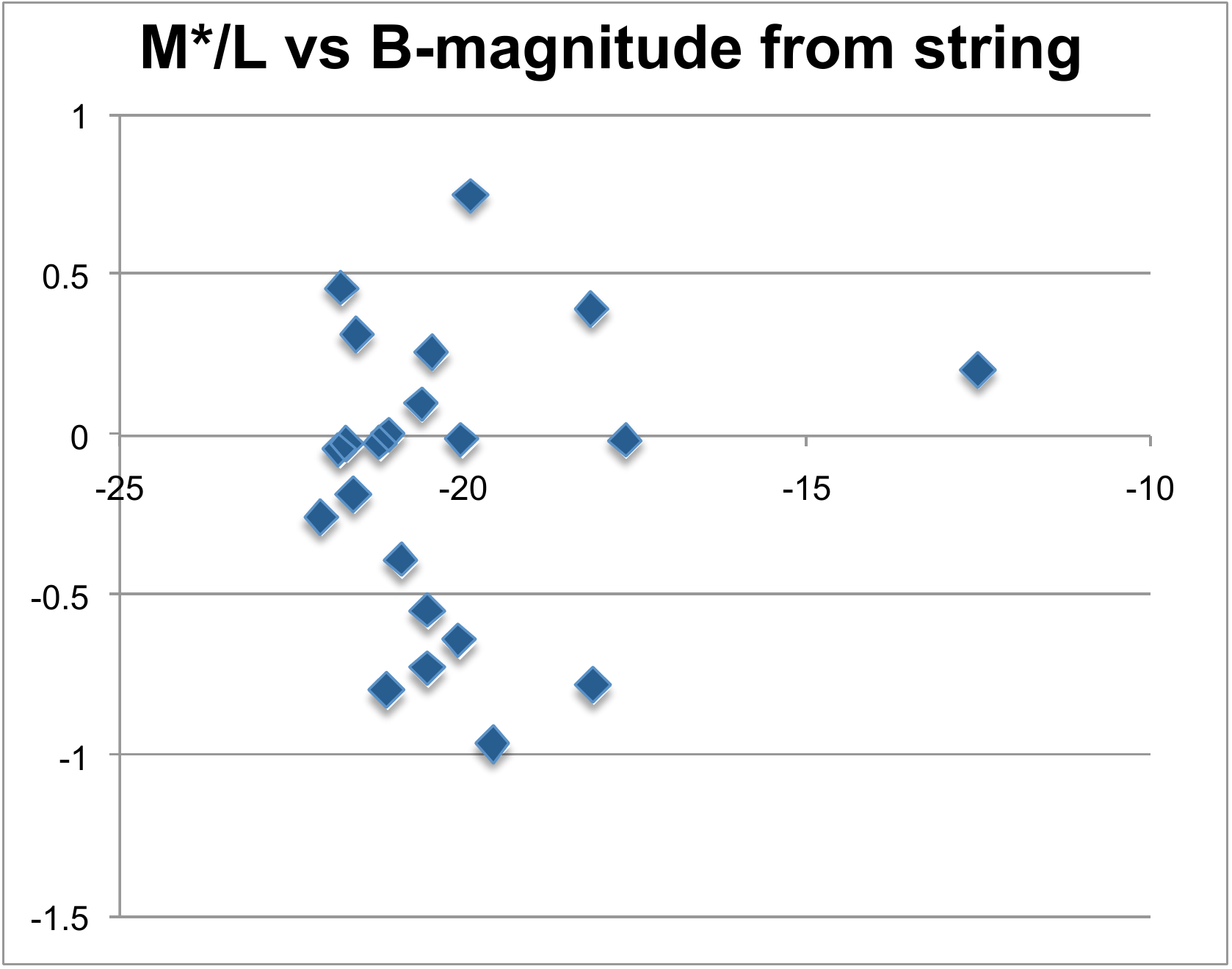}
%\end{center}
\caption{\small The mass to light ratios as derived  from our best fit values.}
 \label{fig:mass-to-light}
 %\vspace{-0.1in}
\end{figure}

As a check we compute the mass to light ratios for the twenty-two galaxies.  We expect the mass-to-light ratios of these spiral galaxies to be smaller than the mass-to-light ratio of the sun which is set to unity~\footnote{We thank Frank van den Bosch for useful discussion regarding this point.}.  The reason being that these galaxies have many  young stars which are very bright. Compare to the sun they emit more light for the same mass.  We would therefore like to see that the mass-to-light ratios of our galaxies to lie between $1< \Upsilon_{galaxy} <0$.   The results are tabulated in the last column of  Table~\ref{table:bestfit} and shown in Figure~\ref{fig:mass-to-light}.  We compare only the B-band magnitude of these galaxies to the B-band magnitude of the sun, $M_{B} = 5.48$.  Ideally one would like to compare the absolute magnitude in the K-band, which is believed to be a better   indicator of the stellar mass.  Unfortunately in this data set the K-band data is not available.  

\begin{figure}[h!!]
%\begin{center}
    %\includegraphics[height=2.0in]{Omega-Upsilon.pdf}
     \includegraphics[height=2.0in]{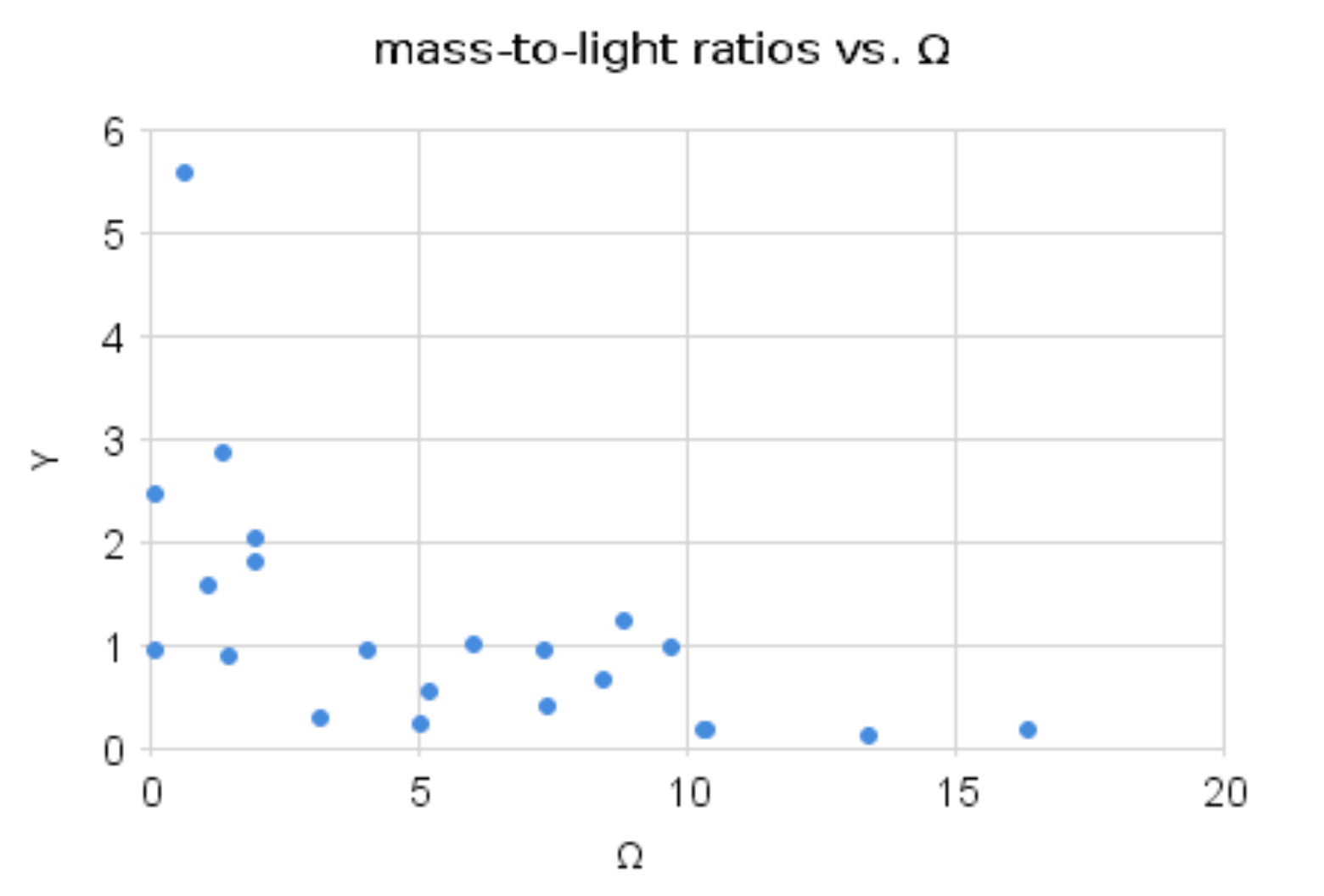}
%\end{center}
\caption{\small The inverse relation between the strength of the background gauge field and the mass-to-light ratios of the galaxies.}
 \label{fig:Upsilon-Omega}
 %\vspace{-0.1in}
\end{figure}

Let us remark that for those galaxies enjoying the benefit of the background rotation the range of mass to light ratios  obtained from our model lies in a much narrower range than that obtained from Dark Matter models(See, for example, \cite{vandenBosch:1999ka}).  In a companion paper~\cite{Cheung:2008fz} we compare our model with the dark matter model by fitting the same group of twenty-two galaxies with Navarro-Frenk-White~\cite{Navarro:1996gj} profile.  Indeed the values of the mass-to-light ratio span less than two orders of magnitude while those from the NFW profile for these twenty-two galaxies scatter five orders of magnitude.   The total mass of the galaxies computed from our theory  as well as their mass-to-light ratios are tabulated in Table~\ref{table:bestfit}.   We also notice an intriguing  trade off between  values of the mass-to-light ratio with the strength of the gauge field, shown in Figure~\ref{fig:Upsilon-Omega}.   This relationship deserves further investigation.     Some of the  galaxies have  strictly flat rotation curves beyond the  stellar disks.  MOND~\cite{Milgrom:1983pn, Sanders:2002pf}  can perhaps do a  good  job  fitting them.  However those with linearly rising speed, for example NGC~2403, are difficult for MOND to capture.

\section{Discussion}

In this note we entertain the idea that a string background field contributes a Lorentz-like  force 
to sustain the galactic rotation in spiral galaxies.  
This  force  gives rise to  an  extra centripetal acceleration in addition to the gravitational 
attraction due to the {\em{visible}} stellar mass.  
If not accounted for properly it would appear to be  mass mysteriously missing in the galaxies.  
This  missing mass  is currently accounted for by postulating the existence of  ``Dark Matter.''   
Together with ``Dark Energy,''  they present two important conundrums for physics in the twenty first century.
Until these two roadblocks are removed we cannot arrive at a comprehensive physical theory of  Nature.   
Until hard proof is presented for the nature of Dark Energy and Dark Matter, one should explore all venues for their explanation.  

At this point it is worth mentioning that a critical  reanalysis of  available data on velocity dispersion of F-dwarfs and K-giants in the solar  neighbourhood    by Kuijken and Gilmore concluded that  the data  provided no robust evidence for the existence of any missing mass associated with the galactic disc in the neighbourhood of the Sun~\cite{gilmore}.   Instead a local volume density of  $\rho_0= 0.10M_{sun} pc^{-3}$ is favoured,  which agrees with the value obtained  by star counting.  Dark matter--if existed--would have to exist outside the galactic disk  in the form of  a gigantic halo. 
Their pioneer work was later corroborated  by~\cite{others} using  other sets of A-star, F-star and  G-giant data.  
Note that this observation can be nicely explained by our model as the field only affects the centripetal motion on the galactic plane  but  does not affect the motion perpendicular to the  plane.

With a three parameter phenomenological model   we are able to explain the shape of the rotation curves  of the spiral galaxies way beyond the stellar disks where the majority of the visible matter resides.  Abstractly speaking to capture a ``Universal Rotation Curve''~\cite{URC} one really needs  three parameters to specify the initial slope, where it turns and the final slope.  In this sense our model utilizes just right number of free parameters.   Working from first principles we are formulating a string sigma model description giving rise to this Lorentz force at a macroscopic scale.  We are  also able to explain gravitational lensing phenomenon which is often cited as another strong evidence for the existence of Dark Matter.   However  we are not able to explain the velocity dispersion of galaxies inside a cluster with this simple model.   Some other mechanism will have to  be  invoked to account for it.   We are investigating  if  some other aspects of string theory can naturally give rise  to such a mechanism, but to no avail at the moment of writing.   If succeeded we hope that  other researchers will be encouraged to look harder  for experimental and observational  connections with string theory, however unlikely they may seem at first sight.

\section*{Acknowledgement}
E. C. and K. S. have benefited from useful discussions  with  Frank van den Bosch,  Long-Long Feng,  Chang-De Gong, Jeremy Lim, Tan Lu,  Shude Mao, Wei-Tou Ni,  Fan Wang,  Shuang-Nan Zhang, and Hong-Shi Zong  as well as   with many participants in the Chinese Physical Society  Fall  2006  meeting,  the  7th Sino-German workshop on  ``Galaxies, Super-massive Black Holes and Large Scale Structures,'' and  Chinese High Energy  Physics Society 2006 Annual Meeting.   However the possible mistakes in the paper reflect only the ignorance of the authors.
H.-C. K. is  supported in part by the National Science Foundation and the National Center  for Theoretical Sciences, Taiwan.

\addcontentsline{toc}{section}{Appendix A.  The Galaxy Rotation Curves}
\appendix
\newpage
%\hspace{-1.5in}
\section*{Appendix A.  The Galaxy Rotation Curves}

The  galaxy rotation curves are fit by the three-parameter model.   The stars with error bars indicate the data points 
while the dots indicate theoretical prediction.  

\begin{figure}  [h!]   
\begin{center}
$
\begin{array}{cc}   
 \includegraphics[height=1.8in]{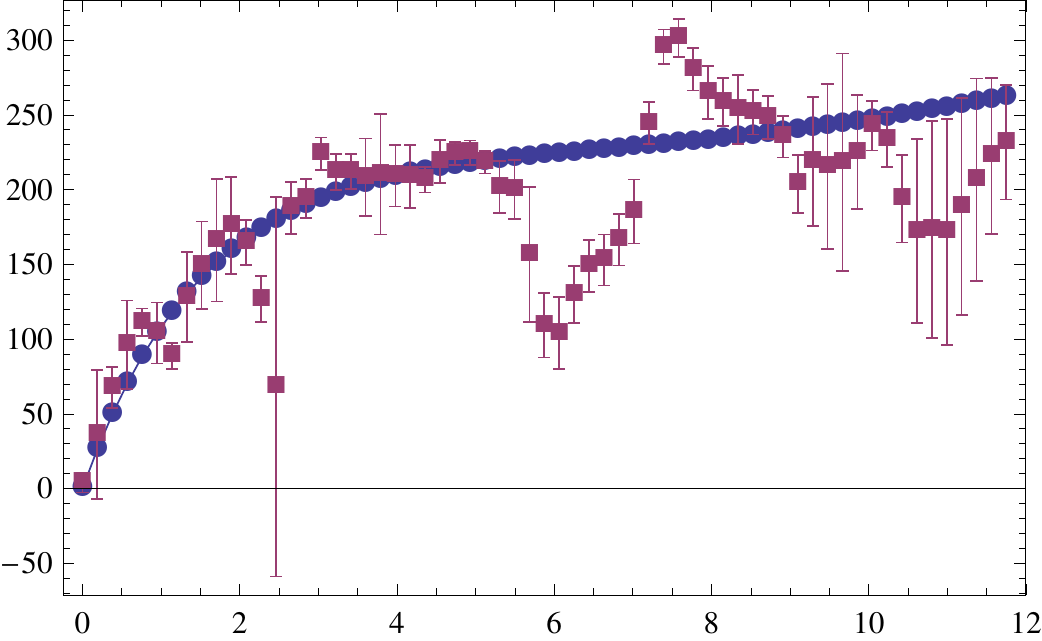} &
\includegraphics[height=1.8in]{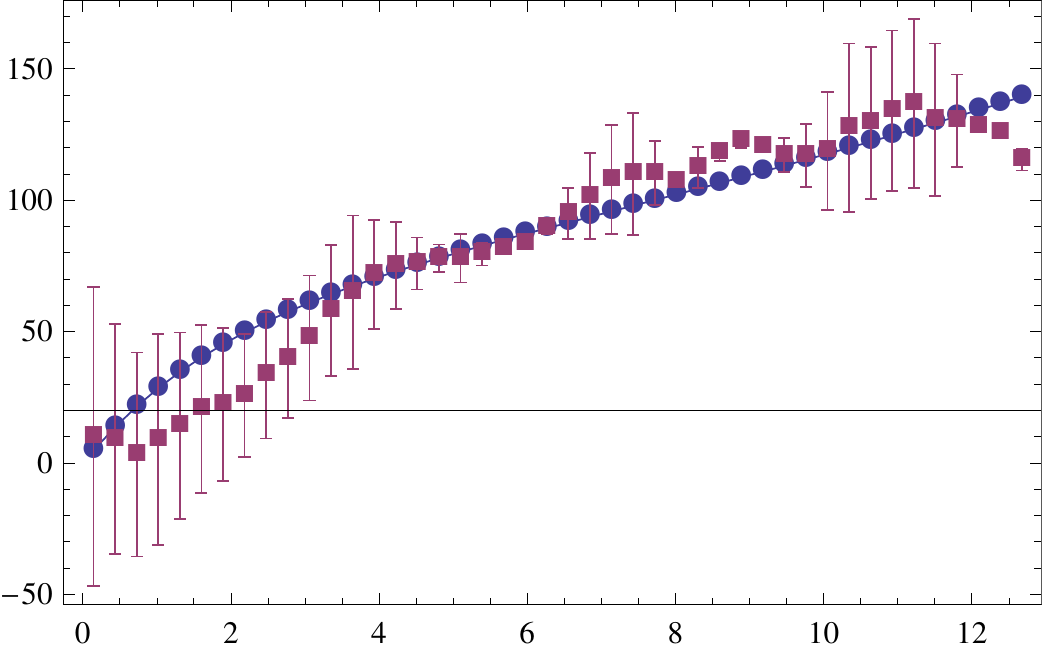}  \\
\mbox{\bf{NGC0628}}& \mbox{\bf{NGC0925}} \\
\includegraphics[height=1.8in]{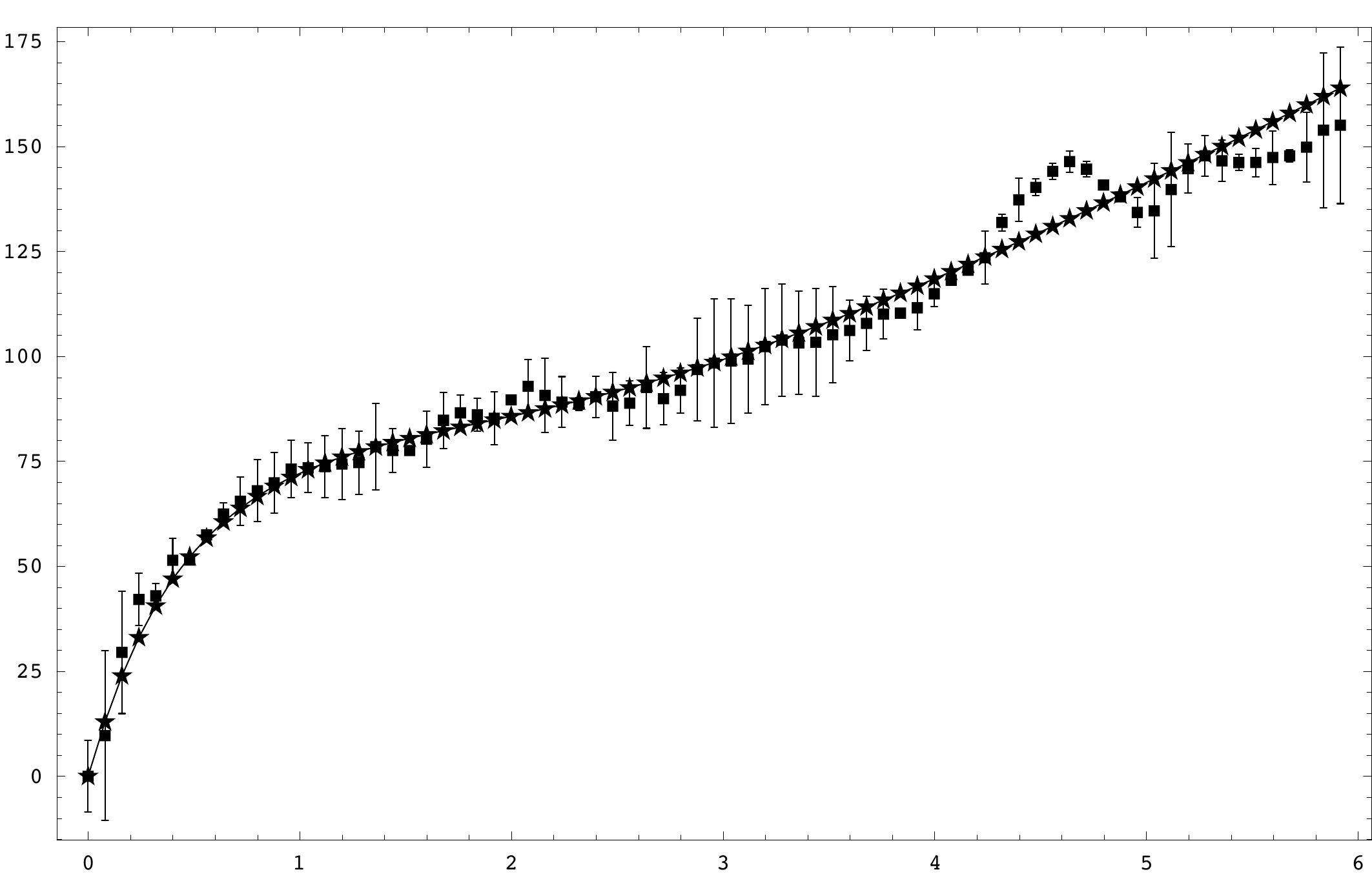} &
\includegraphics[height=1.8in]{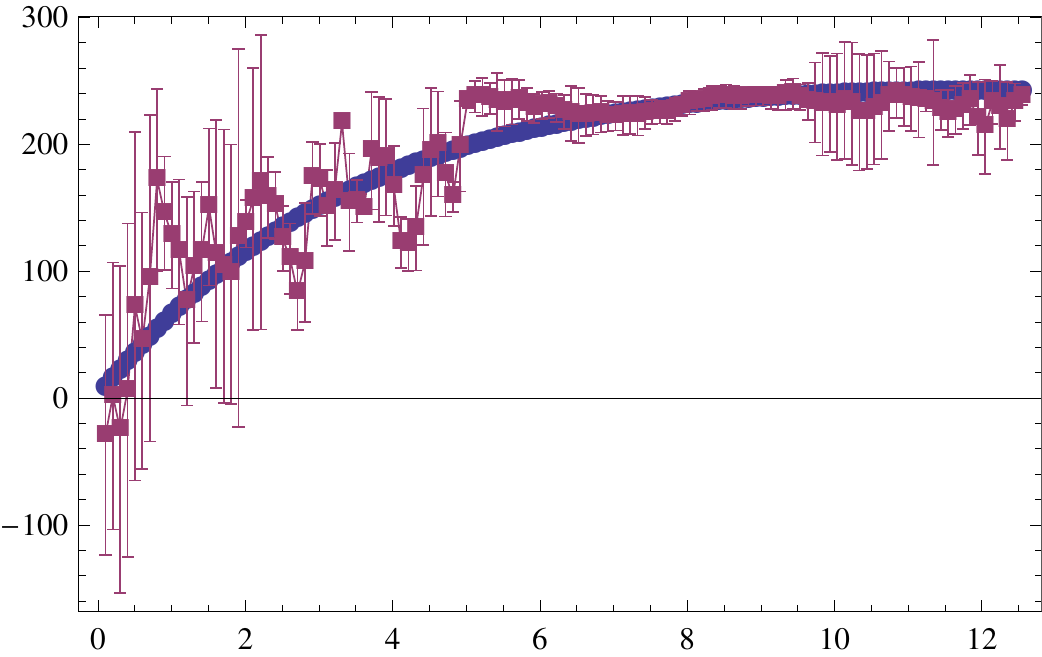}\\
 \mbox{\bf{NGC2403}}& \mbox{\bf{NGC3031}}  \\
%  \includegraphics[height=1.8in]{ngc3184_plot.pdf} &
%\includegraphics[height=1.8in]{ngc3198_plot.pdf} \\
%\mbox{\bf{NGC3184}}& \mbox{\bf{NGC3198}} \\
 \includegraphics[height=1.8in]{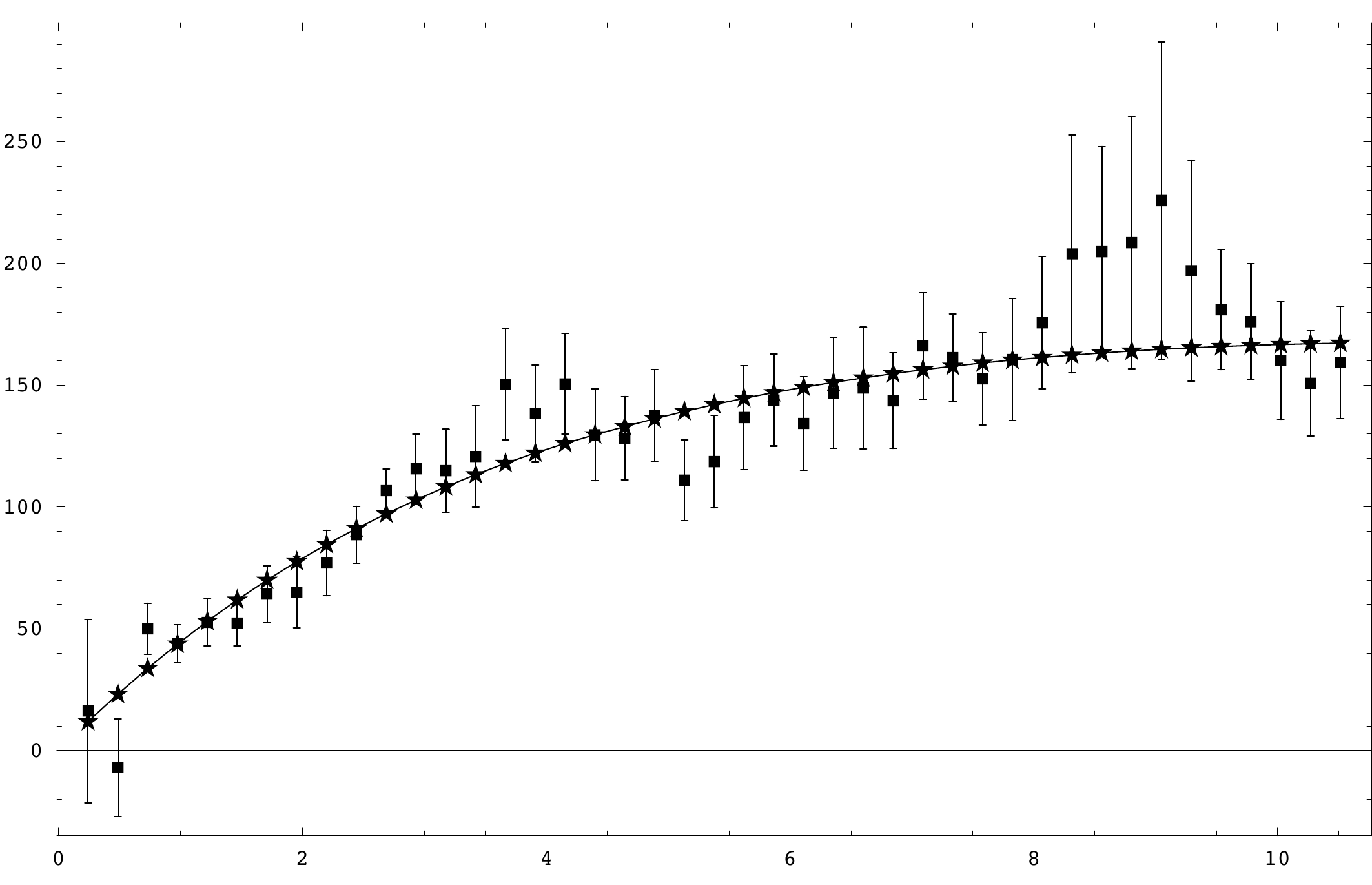} &
\includegraphics[height=1.8in]{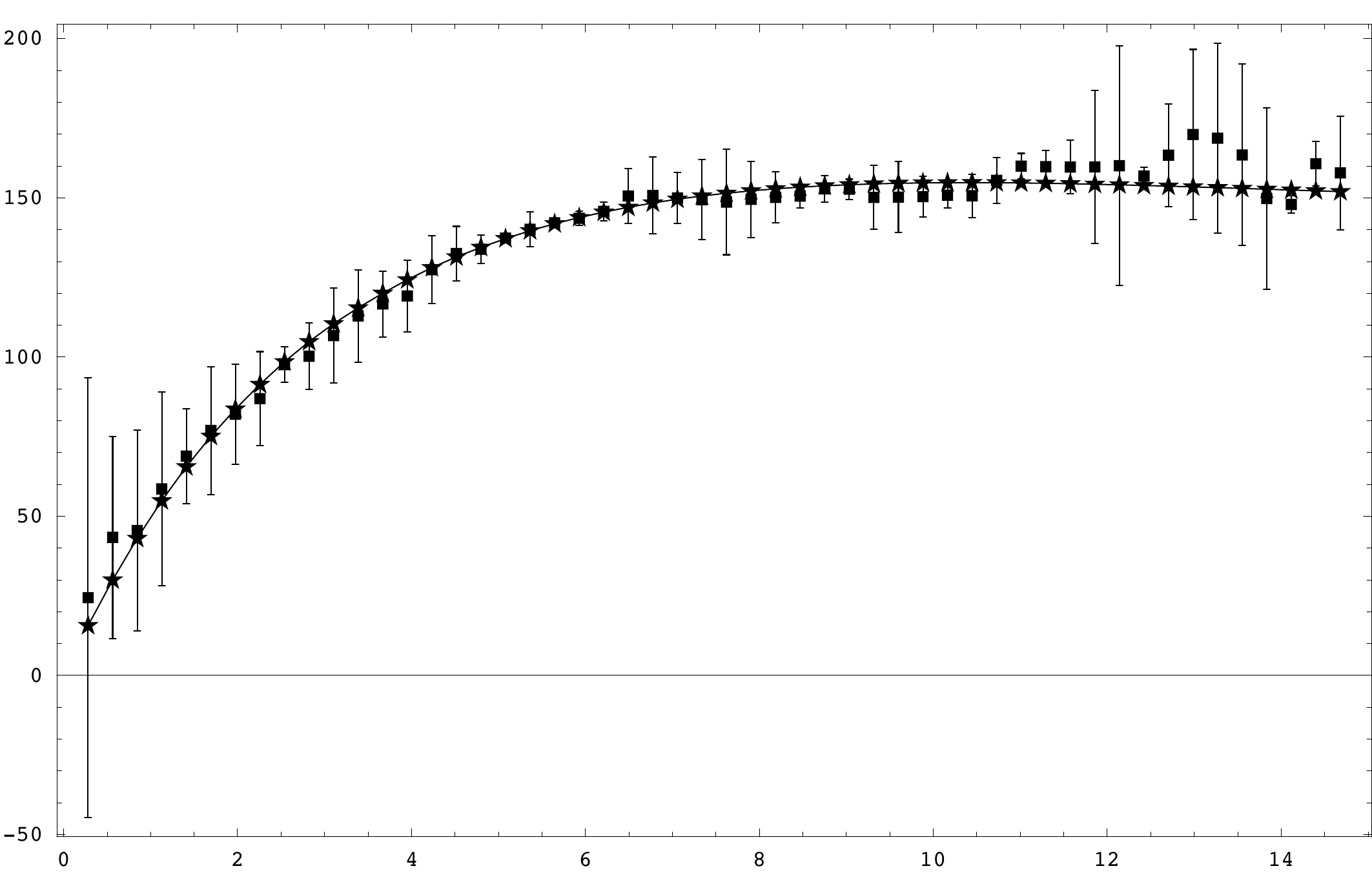} \\
\mbox{\bf{NGC3184}}& \mbox{\bf{NGC3198}} 
  \end{array}
$
\end{center}
\end{figure}

\clearpage

\begin{figure}  [h!]   
\begin{center}
$
\begin{array}{cc}  
%\includegraphics[height=1.8in]{ngc2403_plot.pdf} &
% \includegraphics[height=1.8in]{ngc3031_plot.pdf}\\
% \mbox{\bf{NGC2403}}& \mbox{\bf{NGC3031}}  \\
%\includegraphics[height=1.8in]{ngc3184_plot.eps} &
%\includegraphics[height=1.8in]{ngc3198_plot.eps} \\
%\mbox{\bf{NGC3184}}& \mbox{\bf{NGC3198}} \\
%\mbox{}& \mbox{} \\
% \mbox{}& \mbox{} \\
 \includegraphics[height=1.8in]{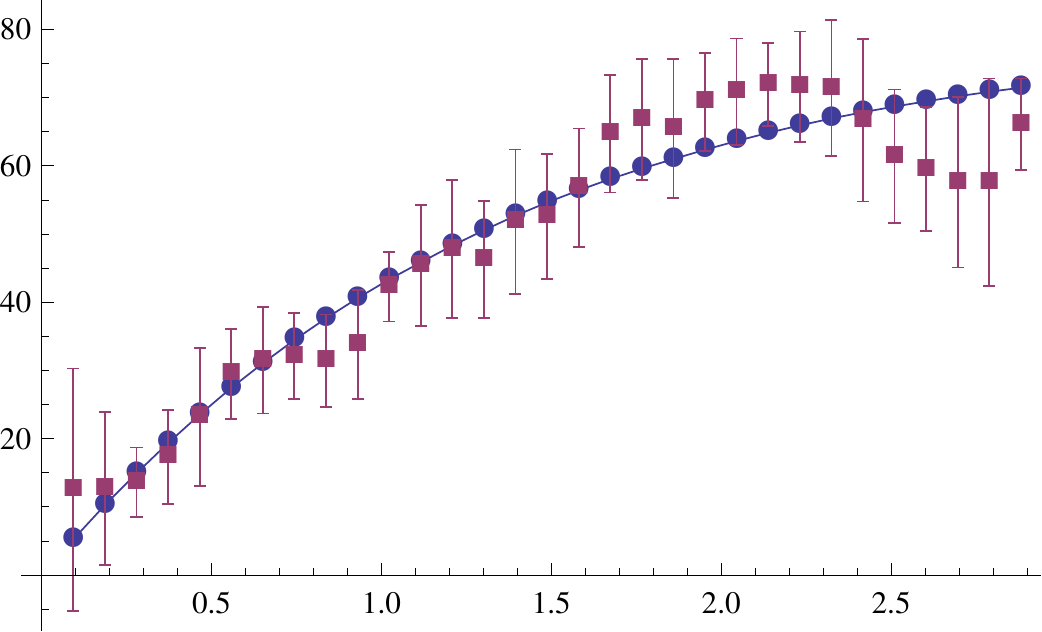} &
  \includegraphics[height=1.8in]{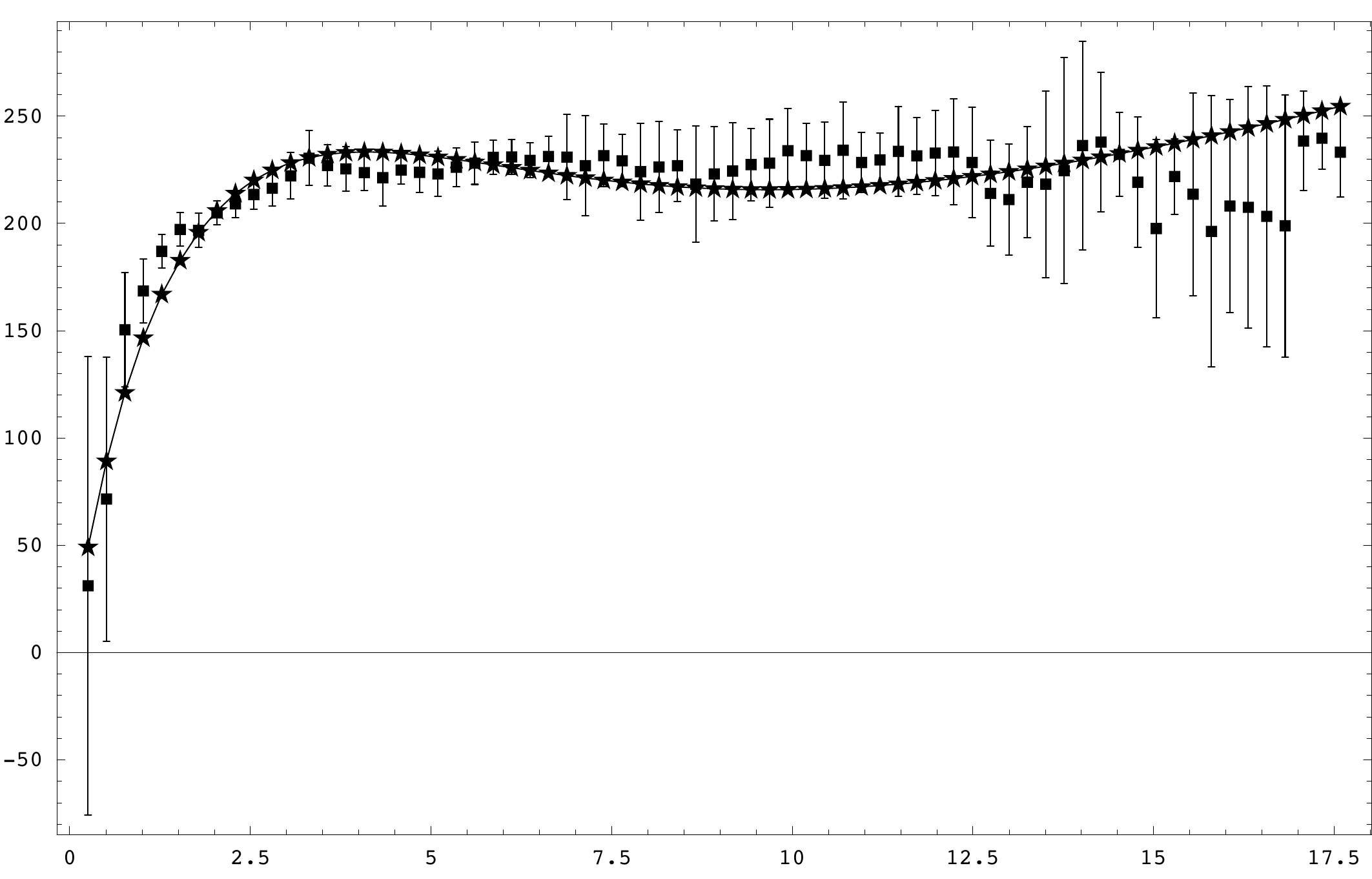} \\
 \mbox{\bf{NGC2976}}& \mbox{\bf{NGC3521}}  \\
\includegraphics[height=1.8in]{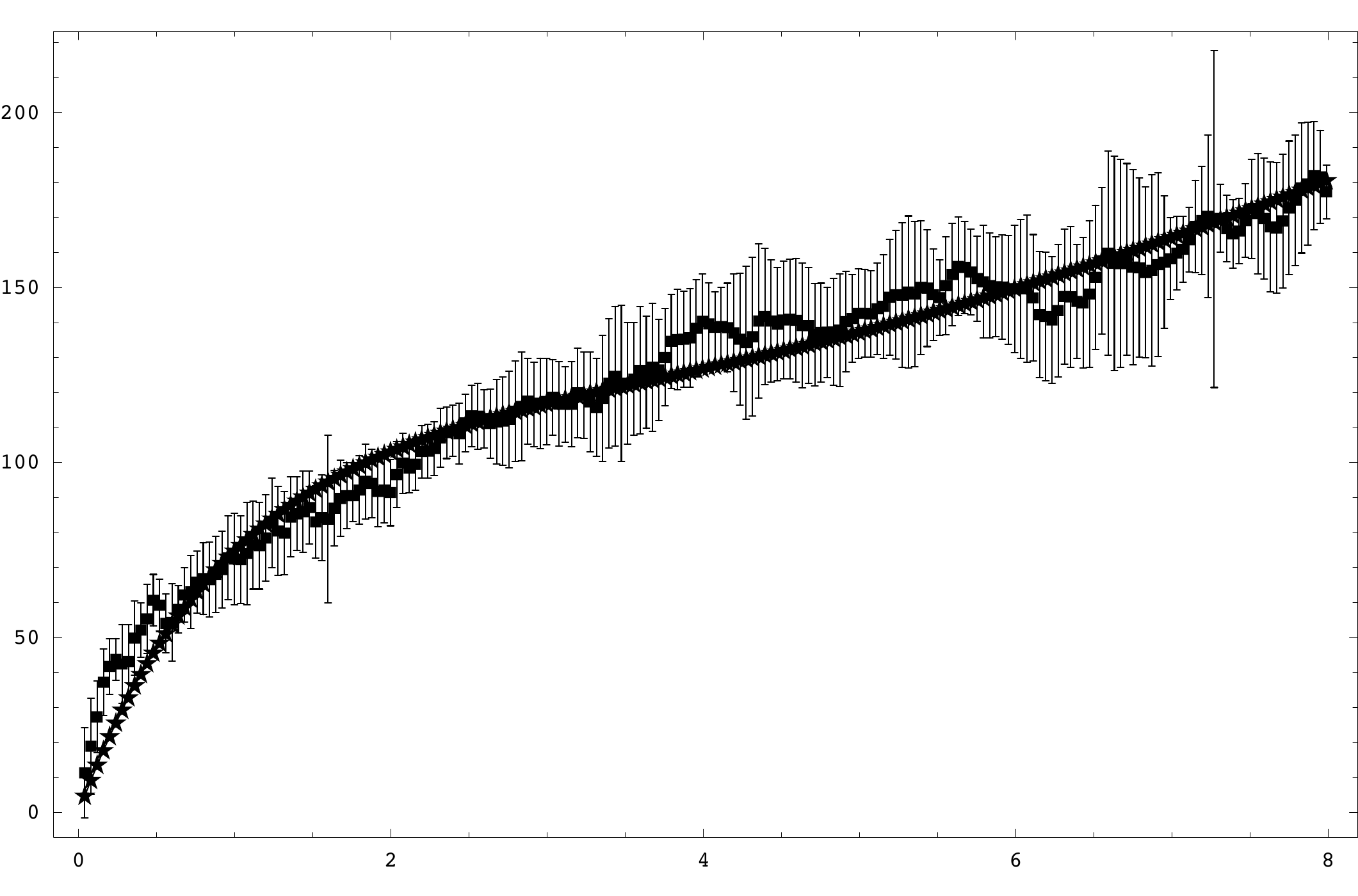} &
 \includegraphics[height=1.8in]{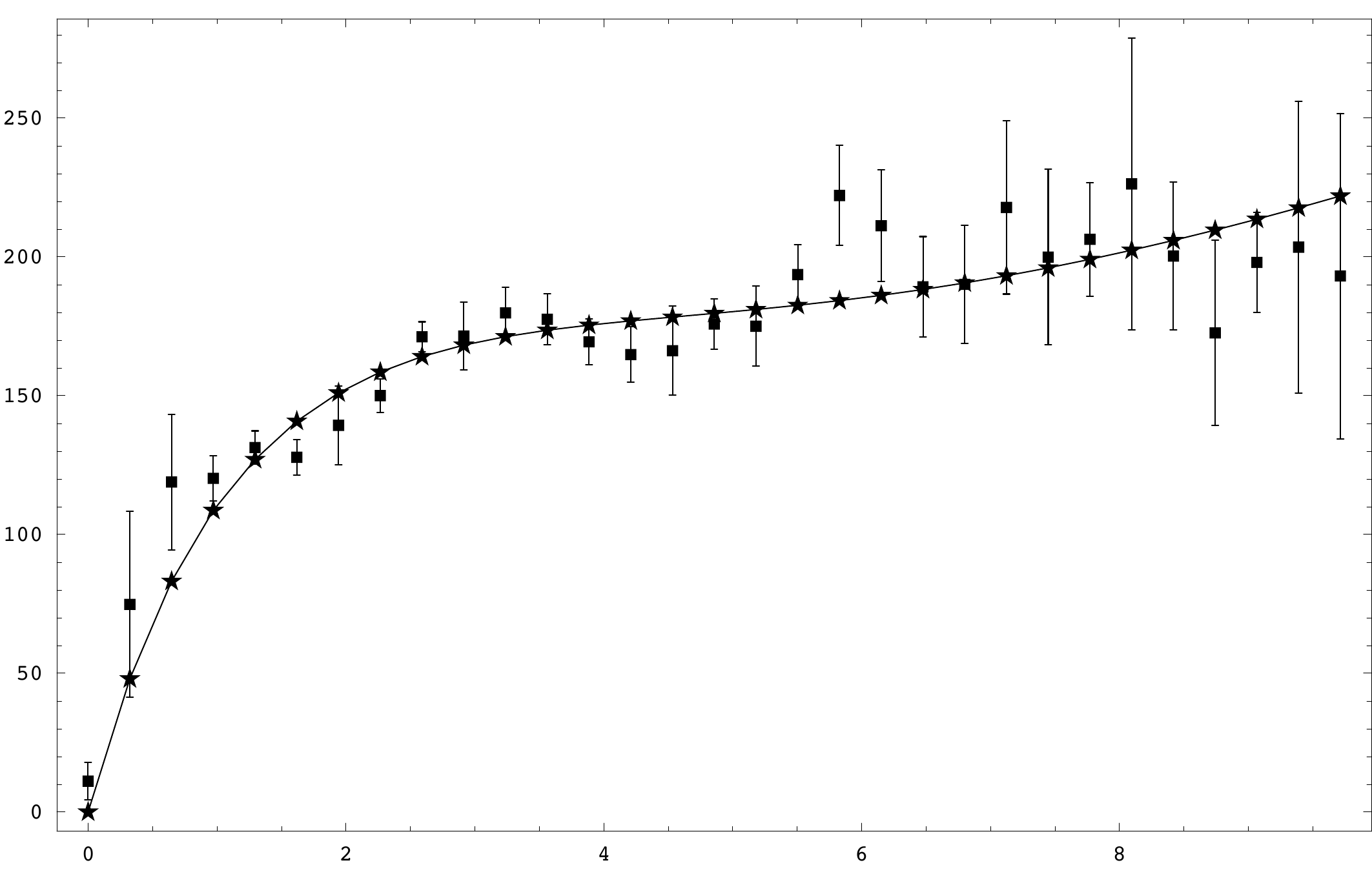}  \\
 \mbox{\bf{NGC3621}}& \mbox{\bf{NGC3938}} \\
  \includegraphics[height=1.8in]{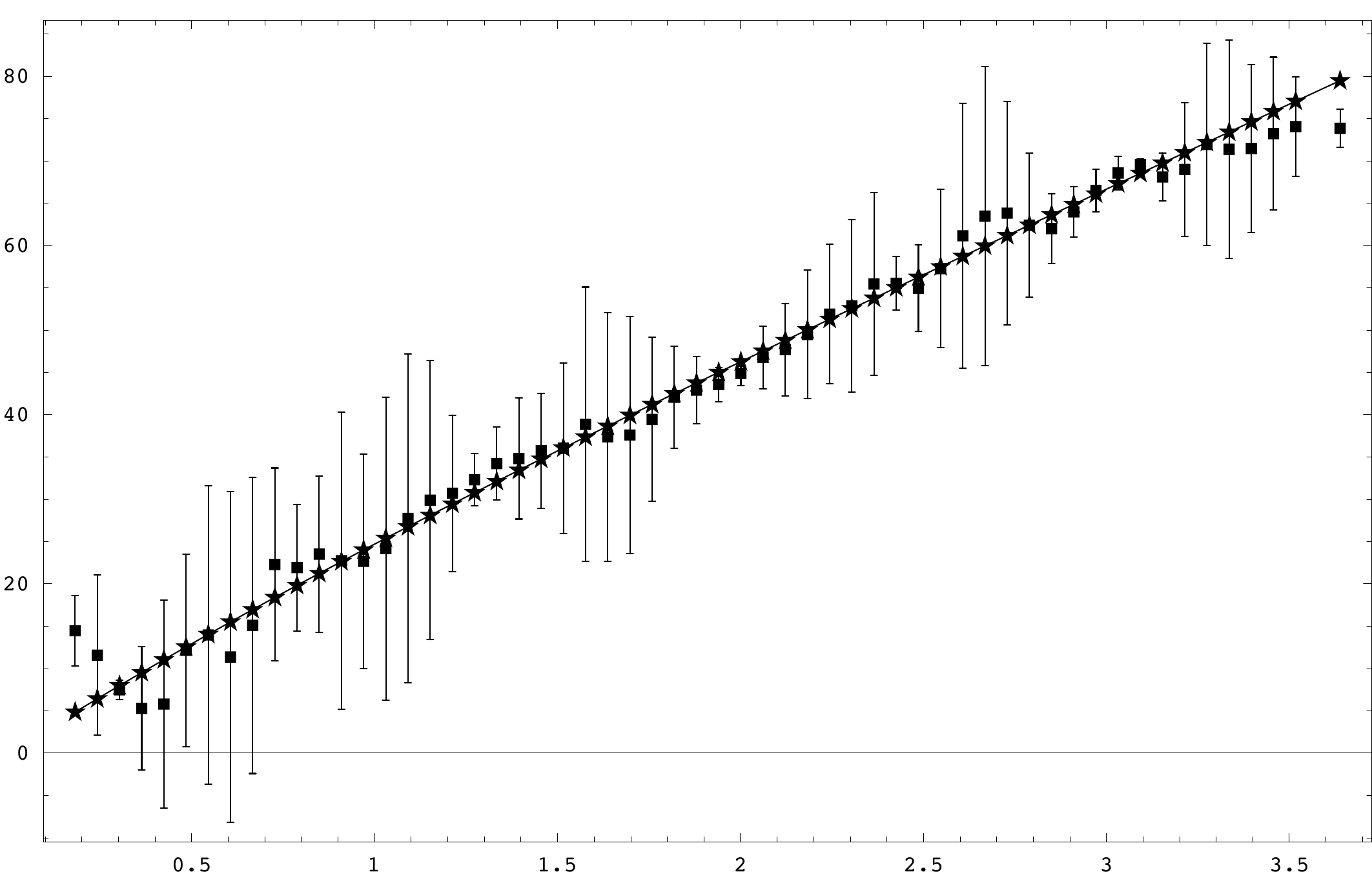} &
  \includegraphics[height=1.8in]{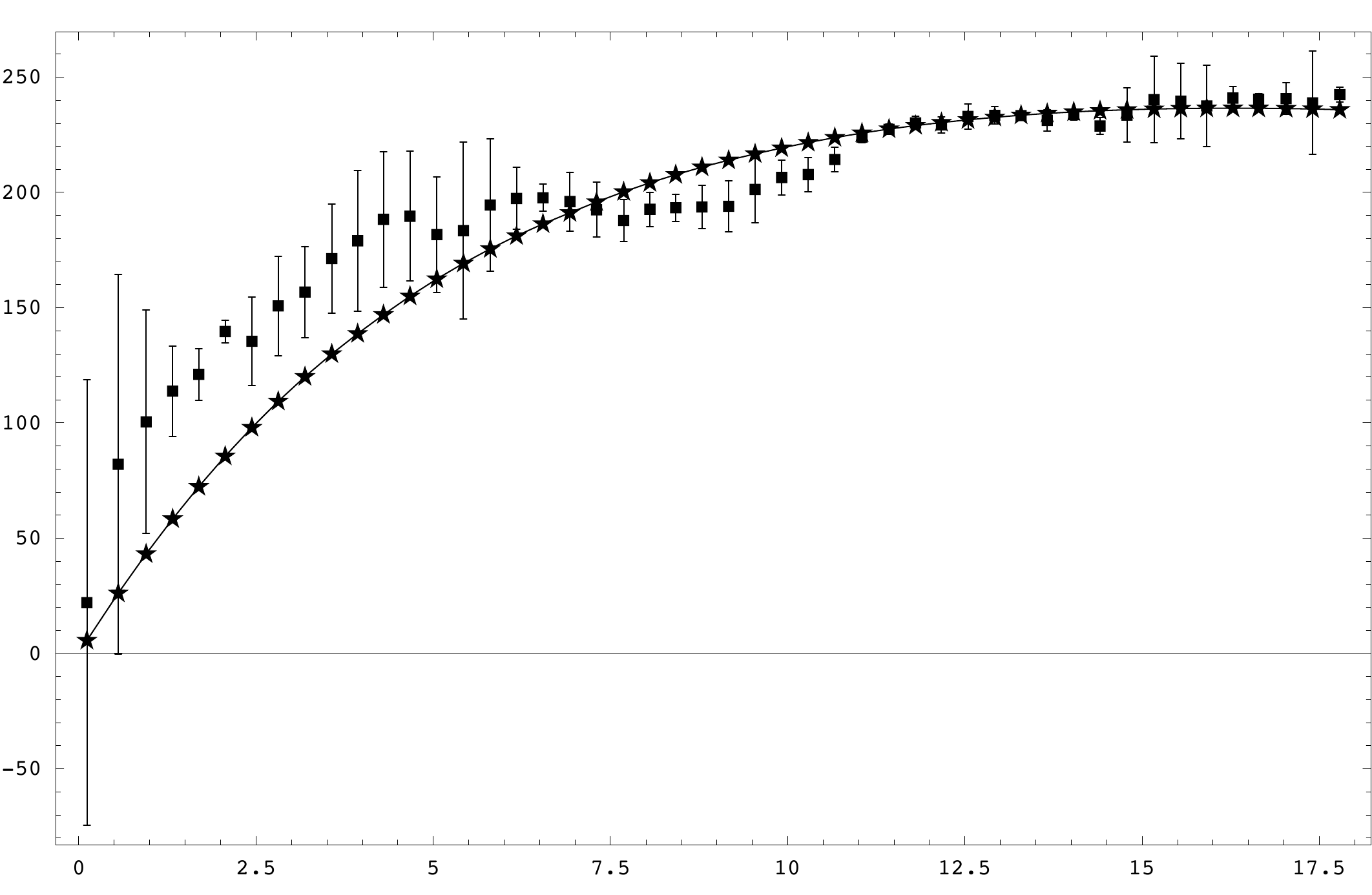} \\
 \mbox{\bf{NGC4236}}& \mbox{\bf{NGC4321}}  \\
%\\ %
 \includegraphics[height=1.8in]{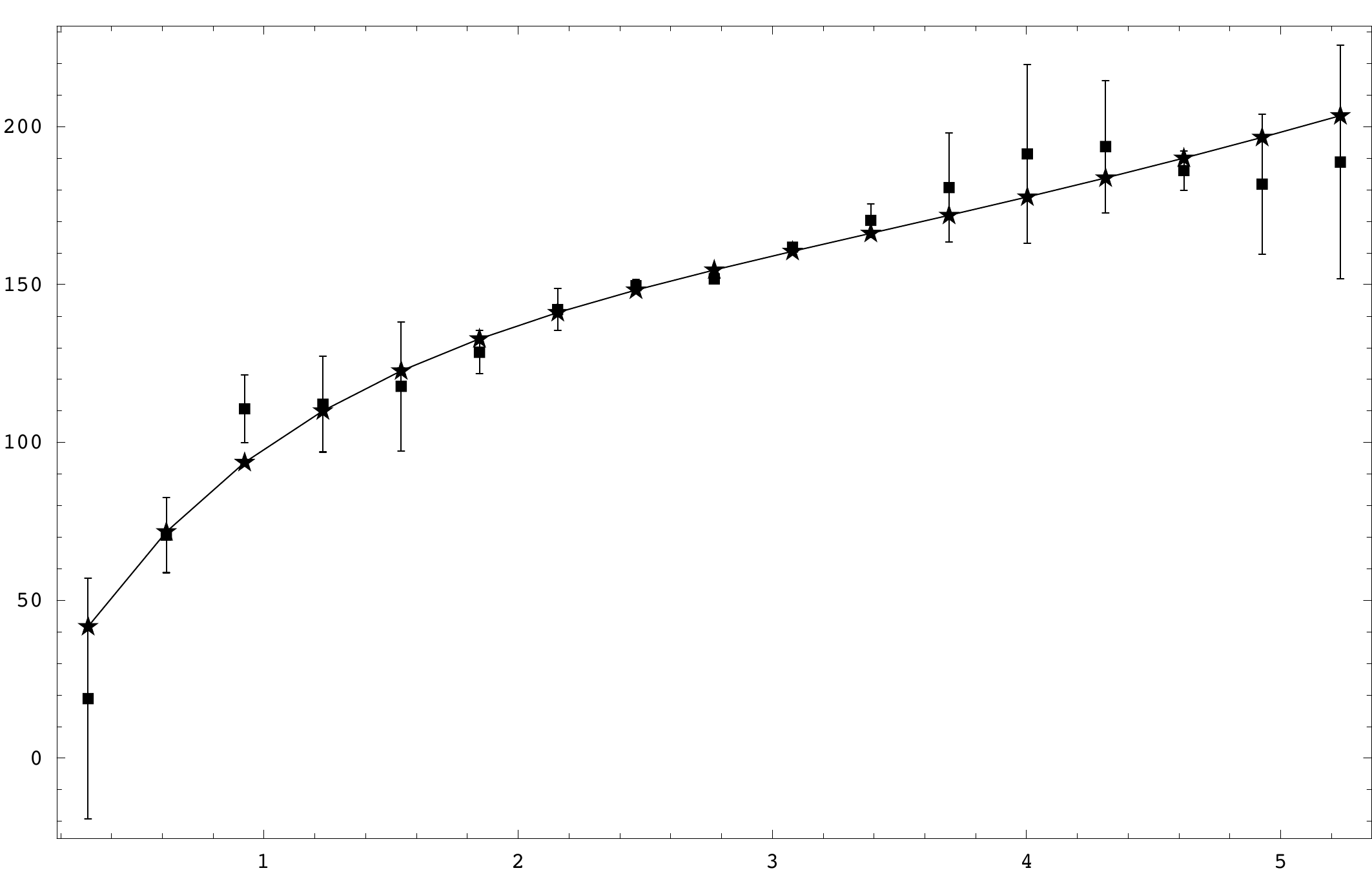} &
 \includegraphics[height=1.8in]{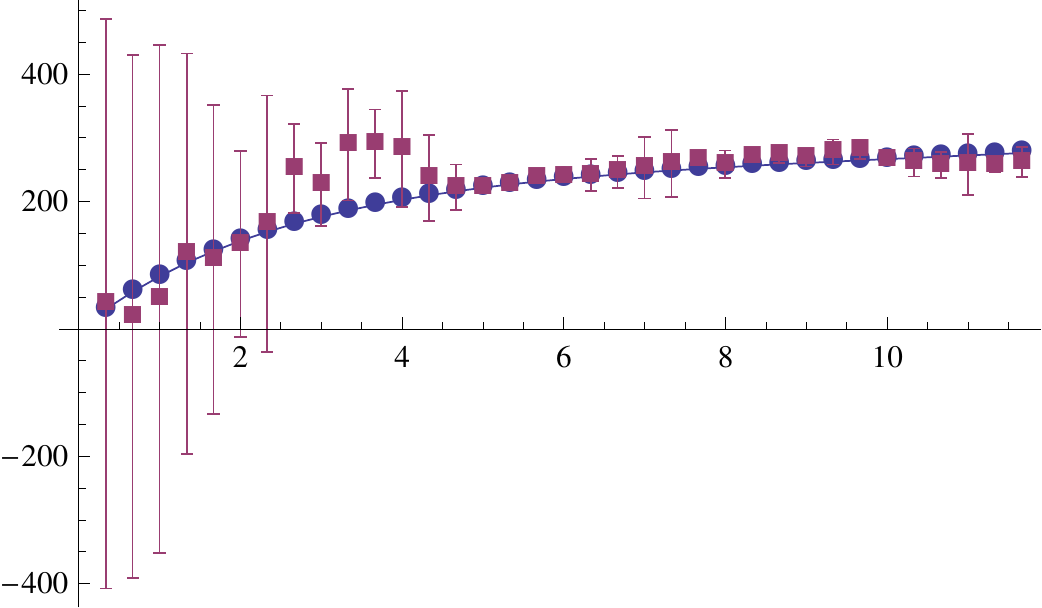}  \\
\mbox{\bf{NGC4569}}& \mbox{\bf{NGC4579}}  
 \end{array} 
$
\end{center}
\end{figure}

\clearpage

\begin{figure}  [h!]   
\begin{center}
$
\begin{array}{cc} 
 %
% \includegraphics[height=1.8in]{ngc4569_plot.pdf} &
% \includegraphics[height=1.8in]{ngc4579_plot.pdf}  \\
 %\mbox{\bf{NGC4569}}& \mbox{\bf{NGC4579}}  \\
%
 \includegraphics[height=1.8in]{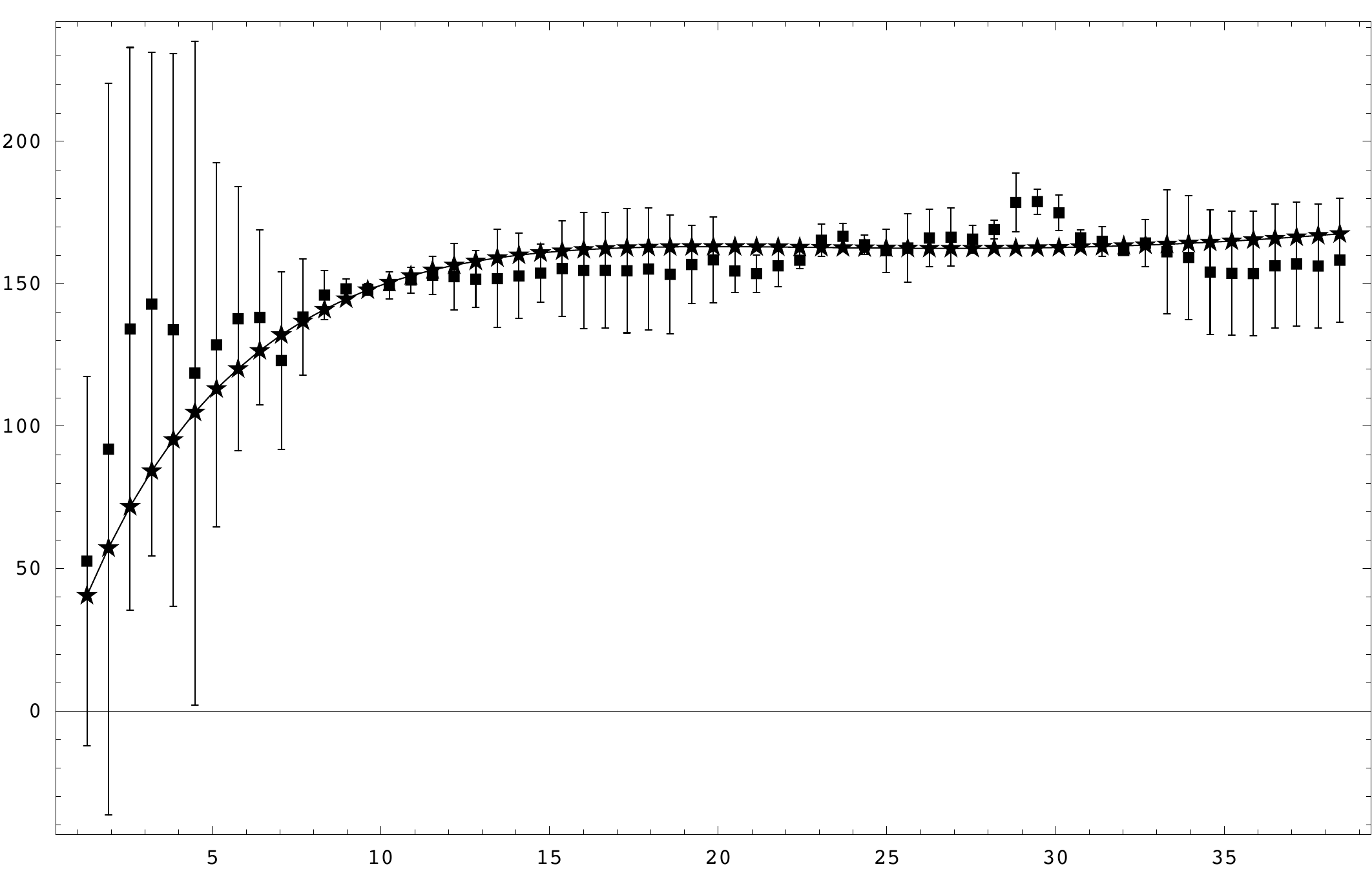} &
  \includegraphics[height=1.8in]{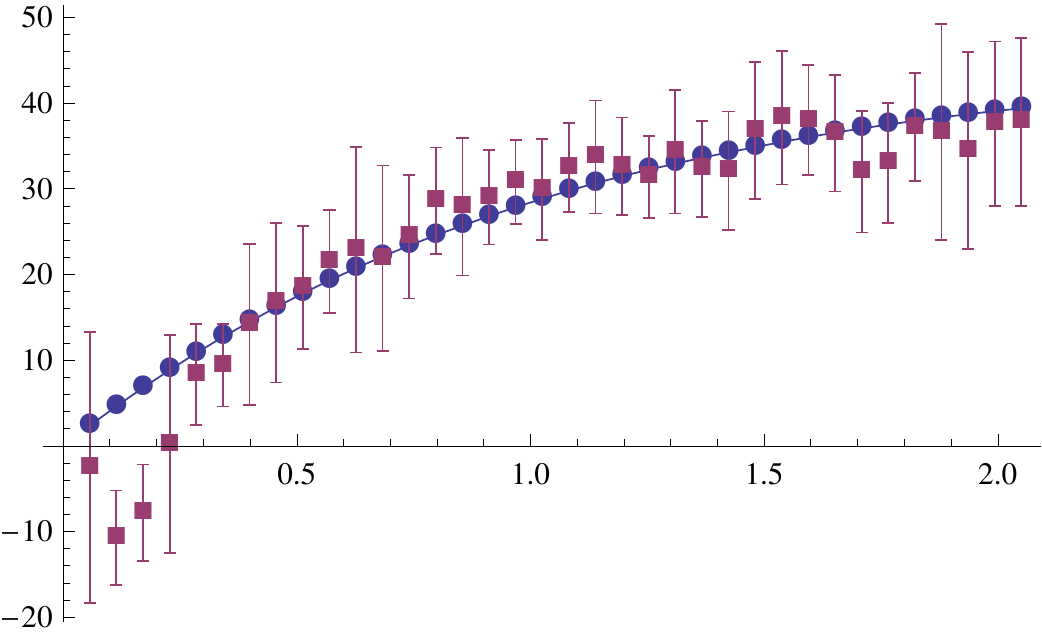}  \\
   \mbox{\bf{NGC4536}} &   \mbox{\bf{NGC4625}}  \\
 \includegraphics[height=1.8in]{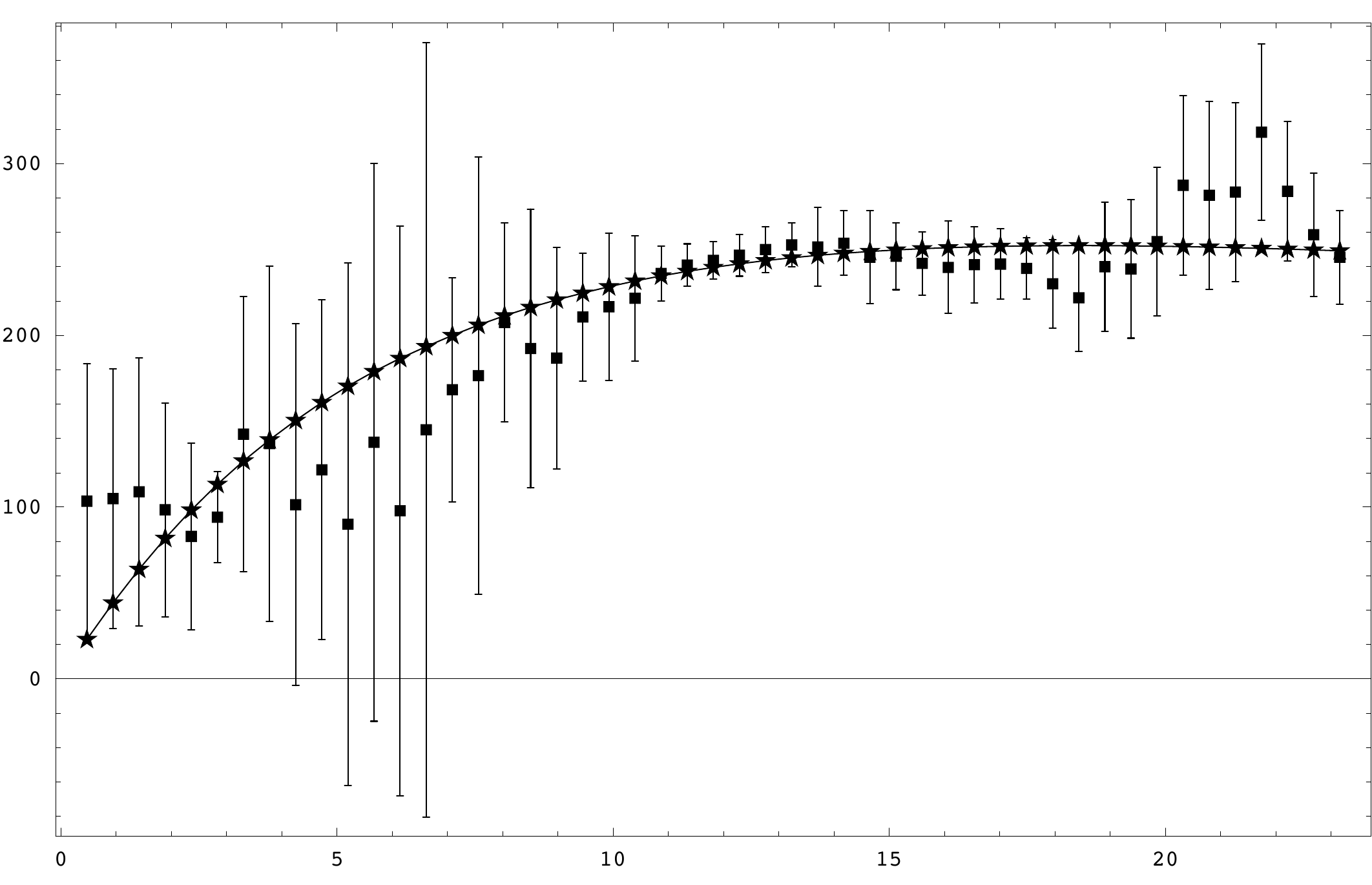}   &
\includegraphics[height=1.8in]{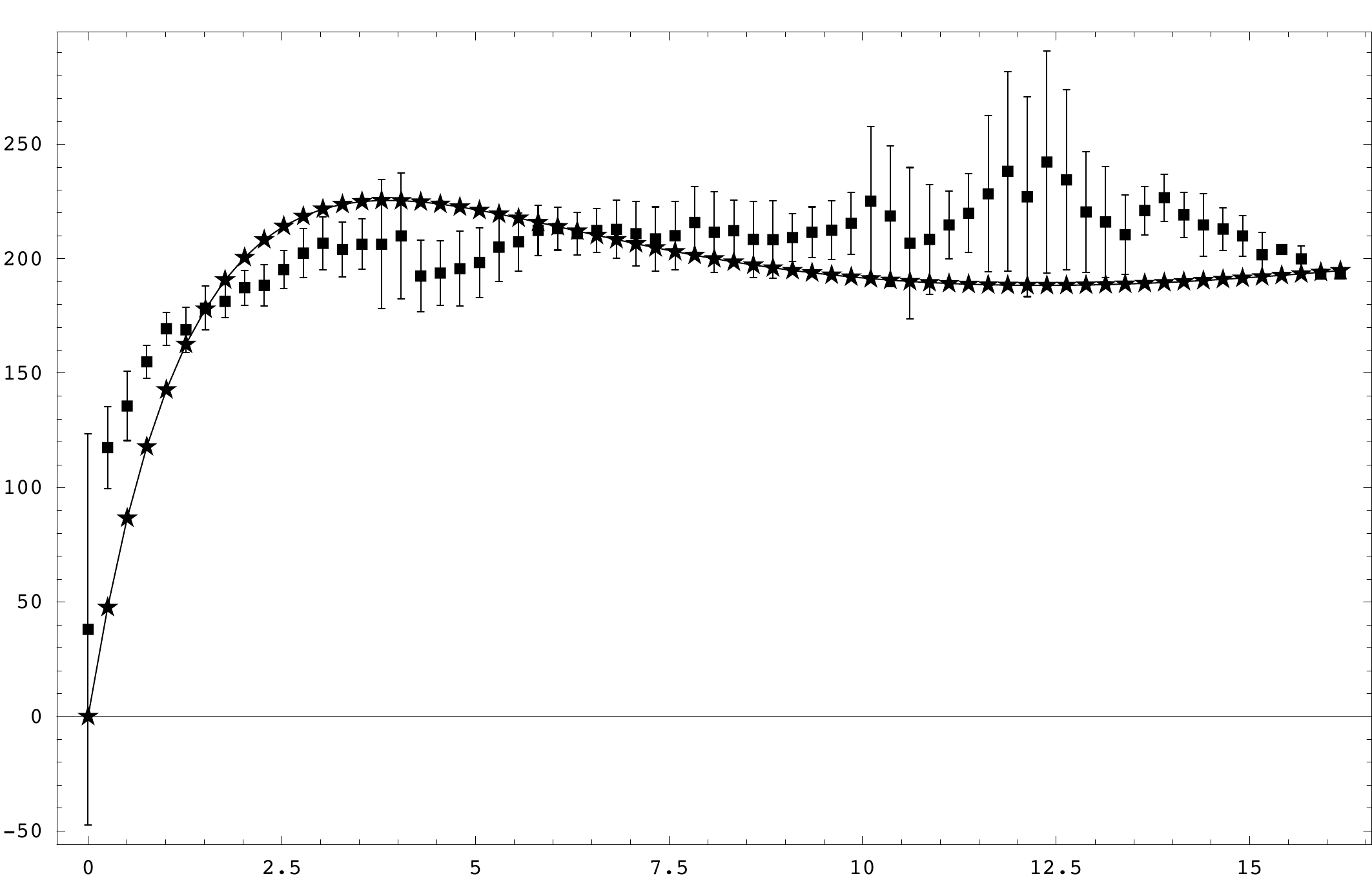} \\
 \mbox{\bf{NGC4725}}& \mbox{\bf{NGC5055}}  \\
% %
  \includegraphics[height=1.8in]{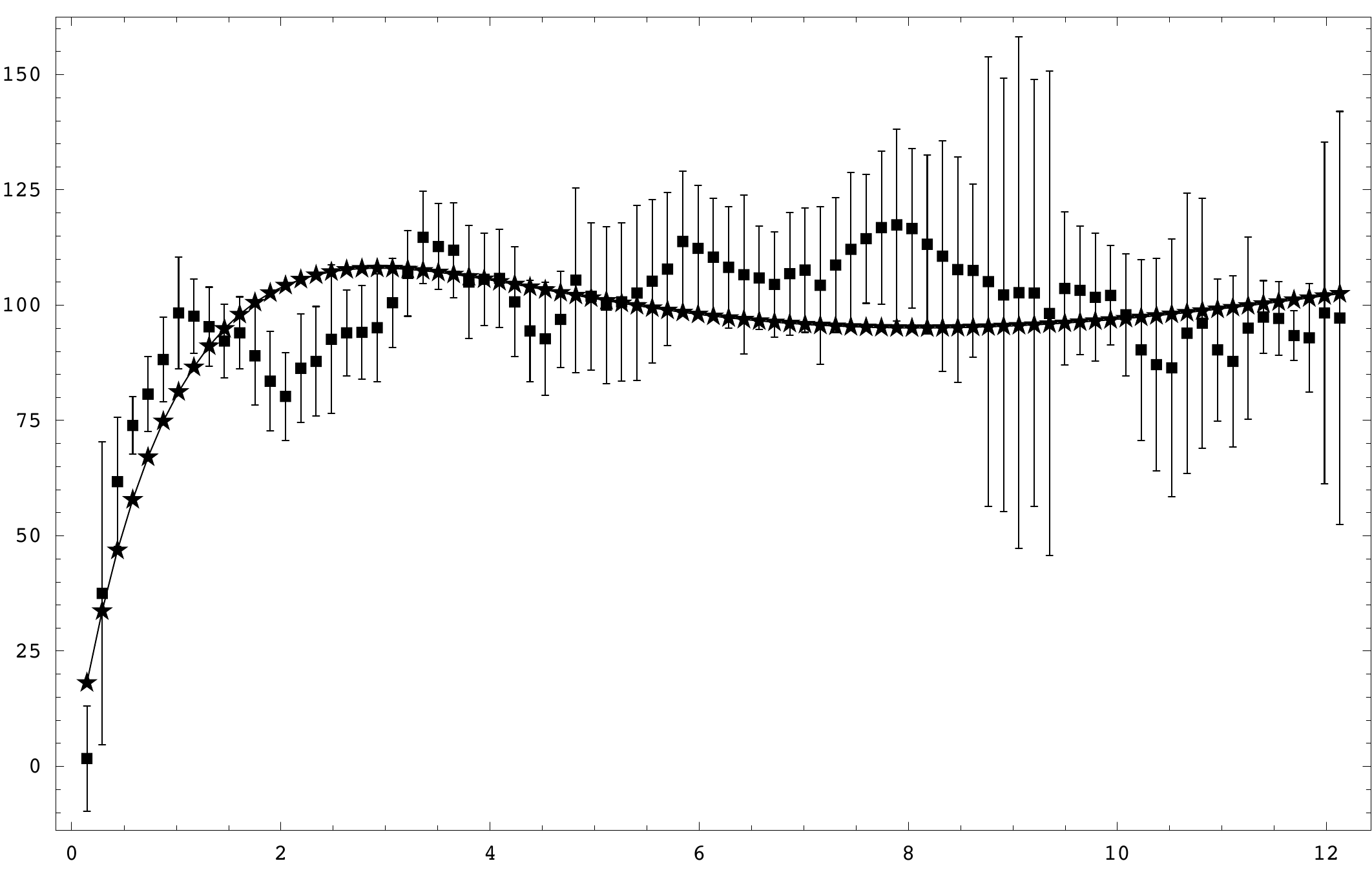} &
   \includegraphics[height=1.8in]{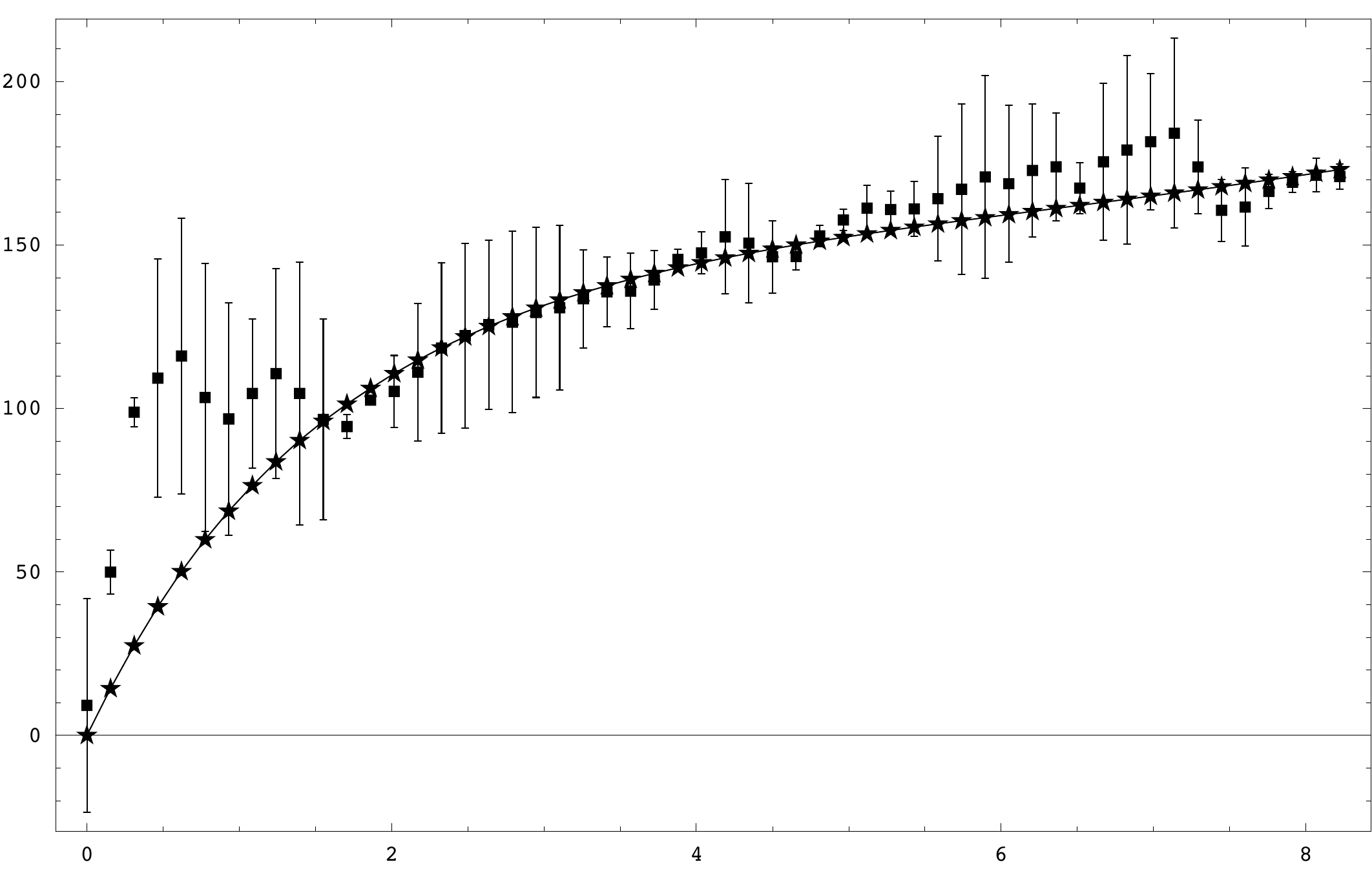} \\
   \mbox{\bf{NGC5194}}& \mbox{\bf{NGC6946}} \\
 \includegraphics[height=1.8in]{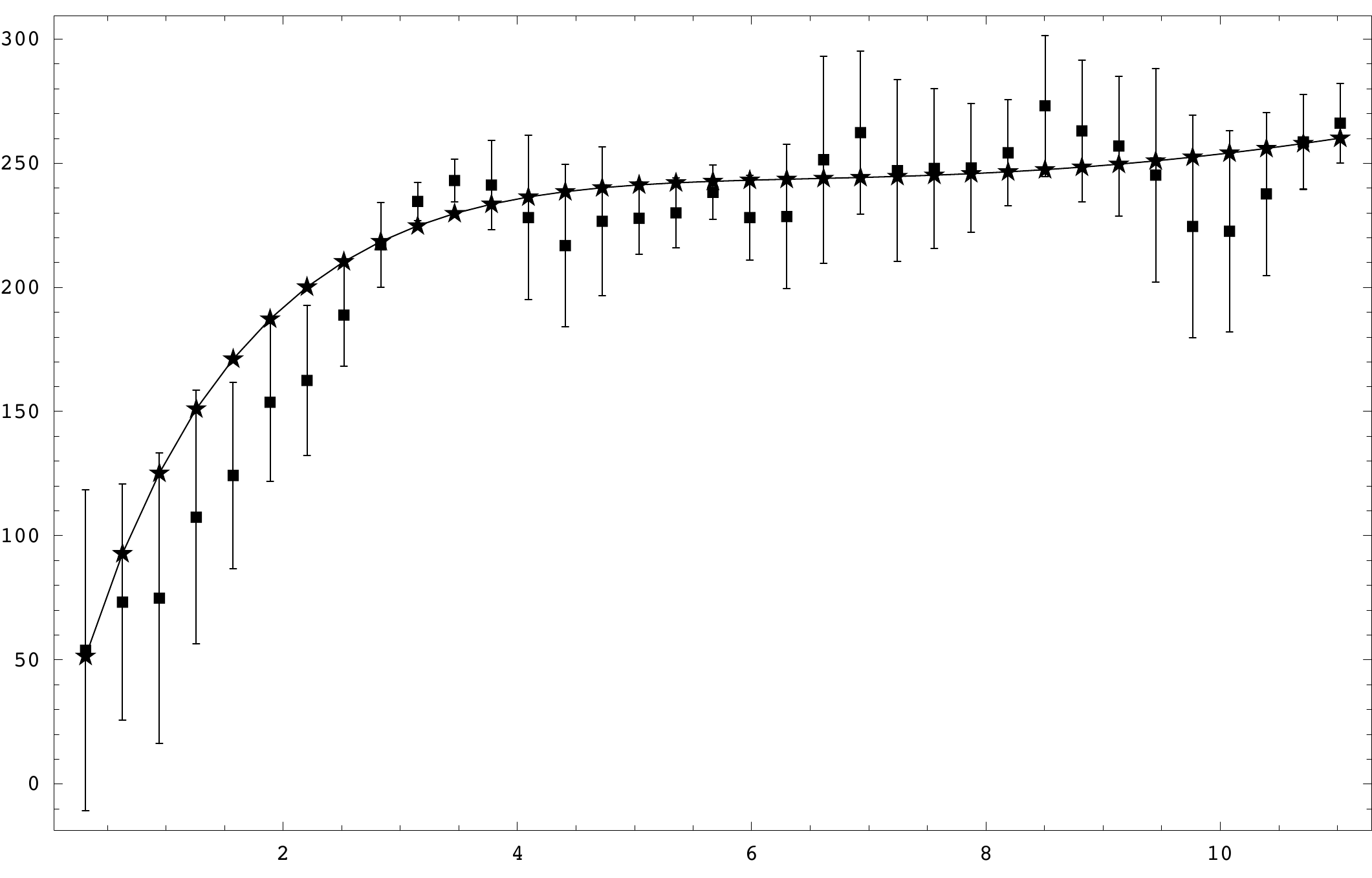} &
  \includegraphics[height=1.8in]{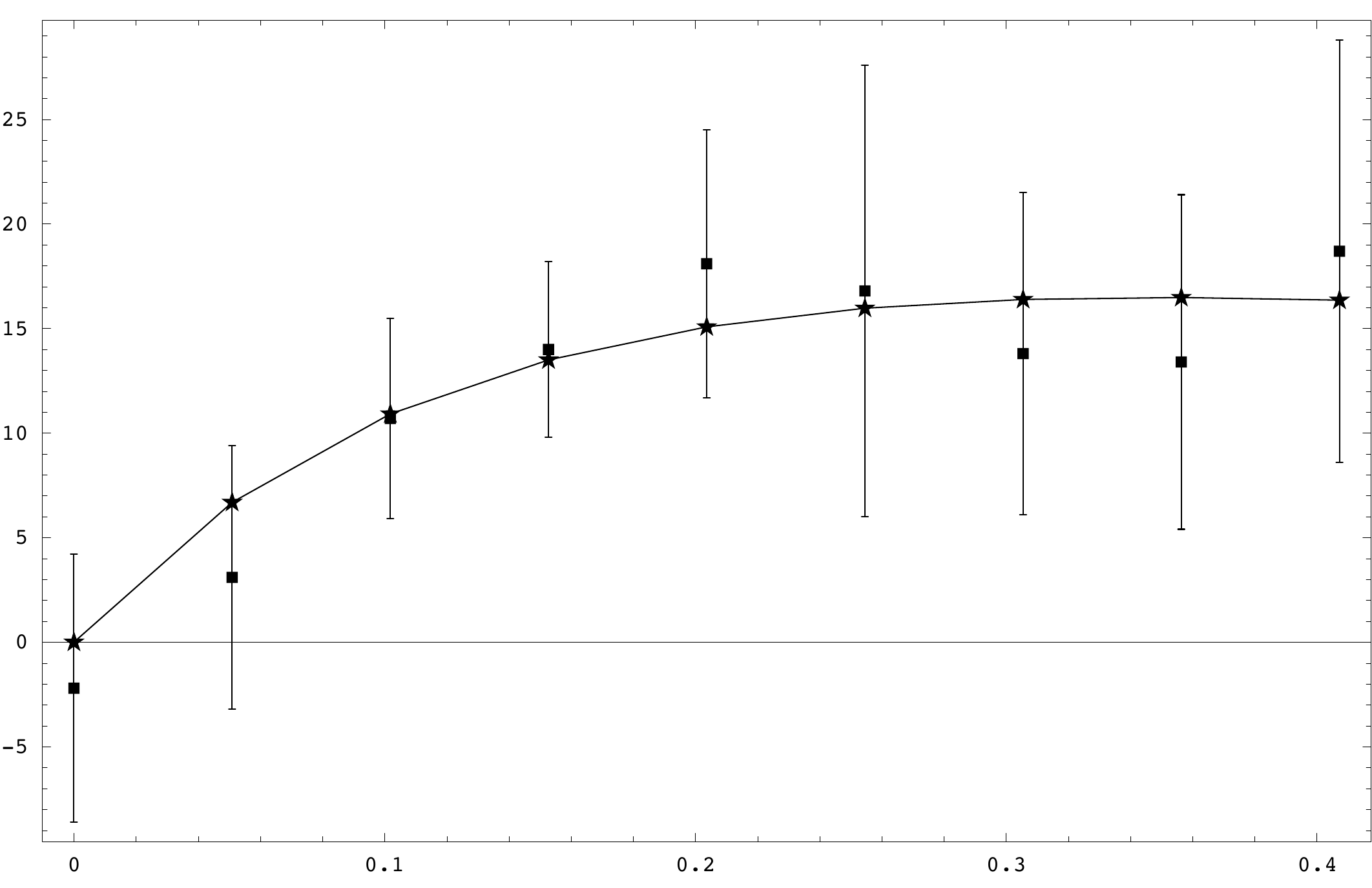} \\
   \mbox{\bf{NGC7331}}& \mbox{\bf{m81dwb}} 
 \end{array} 
$
\end{center}
\end{figure}

%\clearpage

%
%\begin{figure}  [h!]   
%\begin{center}
%$
%\begin{array}{cc} 
% %
%  \includegraphics[height=1.8in]{ngc5194_plot.pdf} &
%   \includegraphics[height=1.8in]{ngc6946_plot.pdf} \\
%    \mbox{\bf{NGC5194}}& \mbox{\bf{NGC6946}} \\
%%
% \includegraphics[height=1.8in]{ngc7331_plot.pdf} &
%   \includegraphics[height=1.8in]{m81dwb_plot.pdf} \\
%    \mbox{\bf{NGC7331}}& \mbox{\bf{m81dwb}} 
% \end{array} 
%$
%\end{center}
%\end{figure}

\end{document}